\documentclass[conference]{IEEEtran}
\IEEEoverridecommandlockouts
\usepackage{url}
\usepackage{gensymb}
\usepackage{cite}
\usepackage{amsmath,amssymb,amsfonts}
\usepackage{algorithmic}
\usepackage{graphicx}
\usepackage{textcomp}
\usepackage{xcolor}
\usepackage{booktabs}
\usepackage{capt-of}
\usepackage{array}
\usepackage{subcaption}
\usepackage{booktabs}
\def\BibTeX{{\rm B\kern-.05em{\sc i\kern-.025em b}\kern-.08em
    T\kern-.1667em\lower.7ex\hbox{E}\kern-.125emX}}

\begin{document}

\title{\huge  Single-Voxel Wireless NeRF for Spatial Spectrum Prediction}
\author{\IEEEauthorblockN{
Raj Bakhunchhe\textsuperscript{1},
Arun Paidimarri\textsuperscript{2}, 
Sara Garcia Sanchez\textsuperscript{2},
Asaf Tzadok\textsuperscript{2},
Ali Tajer\textsuperscript{1}, 
Ish Kumar Jain\textsuperscript{1}}
\IEEEauthorblockA{\textsuperscript{1}\textit{Rensselaer Polytechnic Institute, Troy, NY, USA},\\
\textsuperscript{2}\textit{IBM, Yorktown Heights, NY, USA}
}
}
\newcommand{\AT}[1]{\textcolor{red}{\textbf{AT: }#1}}
\newcommand{\todo}[1]{\textcolor{blue}{\textbf{ToDo: }#1}}

\maketitle

\begin{abstract}
Wireless channel measurements across multiple spatial directions are crucial for AI-driven applications, such as RF digital twins and integrated communication and sensing. However, collecting channel data across large scenes is labor-intensive. 
Wireless NeRFs address this challenge by learning propagation behavior from sparse measurements and synthesizing channel spatial spectrum magnitude at unseen locations. However, existing wireless NeRFs inherit dense volumetric sampling from vision NeRFs, which requires substantial computation. This paper asks whether such dense sampling is necessary for predicting magnitudes of the wireless spatial spectrum. We empirically show that wireless NeRFs are over-parameterized for this task and introduce SV-INGP, a sparse volumetric sampling variant of Instant Neural Graphics Primitives (INGP). Across real-world and simulated datasets, SV-INGP matches the median Structural Similarity Index Measure (SSIM) of the NeRF2 baseline while reducing training time by 184x. These results generalize across LoS and NLoS scenes, sub-6 and millimeter-wave frequencies, and antenna array tapering configurations, suggesting a simpler and more efficient design path for RF digital twins.

\end{abstract}

\section{Introduction}

Emerging AI-based and wireless applications rely on large channel datasets for training machine learning models, evaluating communication algorithms, and building wireless digital twins~\cite{deepmimo,digital_twin_6g,dl_channel_estimation_and_signal_detection_hao_2018,dl_based_channel_estimation_soltani_2019,dl_channel_estimation_beamspace_mmwave_mimo_hengtao_2018}. Applications such as beam management~\cite{ai_ml_for_beam_management,beam_management_for_dense_mmwave_network,two_beams_2021}, integrated sensing and communication~\cite{isac_survey,isac_survey_2026}, channel prediction~\cite{dl_channel_estimation_and_signal_detection_hao_2018,pilot_cavers,pilot_ofdm_van_de_beek}, localization\cite{localization_2015,localization_metaloc_2023}, and AI-driven wireless optimization often 
require realistic directional channel measurement datasets across a wide range of transmitter and receiver locations. 
Existing data curation approaches primarily rely on either simulation or RF measurement. While simulation tools such as Wireless InSite and Sionna Ray Tracing can generate synthetic channels at scale, their accuracy depends heavily on the fidelity of the environment model and material parameters~\cite{sionna_ray_tracing_hyodis,nerf2,voxelrf}. In practice, it is difficult to accurately specify the electromagnetic properties of every object in a real environment, especially in complex indoor deployments with diverse materials, furniture, and clutter. 
On the other hand, collecting large-scale real-world datasets provides high-fidelity channel information but requires extensive measurement campaigns that are time-consuming, labor-intensive, and difficult to scale. Therefore, acquiring such datasets in an accurate yet computationally efficient manner remains a major challenge~\cite{deepmimo,deepsense_6g,csi_benchmark_dataset}.

Recently, Neural Radiance Fields (NeRFs)~\cite{nerf} have emerged as a promising alternative for wireless channel spatial spectrum magnitude prediction. Inspired by the power of NeRFs in computer vision, wireless NeRFs learn a continuous RF scene representation directly from sparse channel measurements at fewer locations in the scene.

Then, the channel spatial spectrum magnitude is subsequently synthesized at previously unseen locations \cite{nerf2,newrf}. Unlike conventional ray tracing approaches, wireless NeRFs do not require explicit knowledge of material properties or detailed environment modeling. Instead, they learn scene-dependent propagation behavior directly from measured data~\cite{nerf2,newrf,nerf-apt,NeRF2-ss,rfcanvas}.

\begin{figure}
    \centering
    \includegraphics[width=\linewidth]{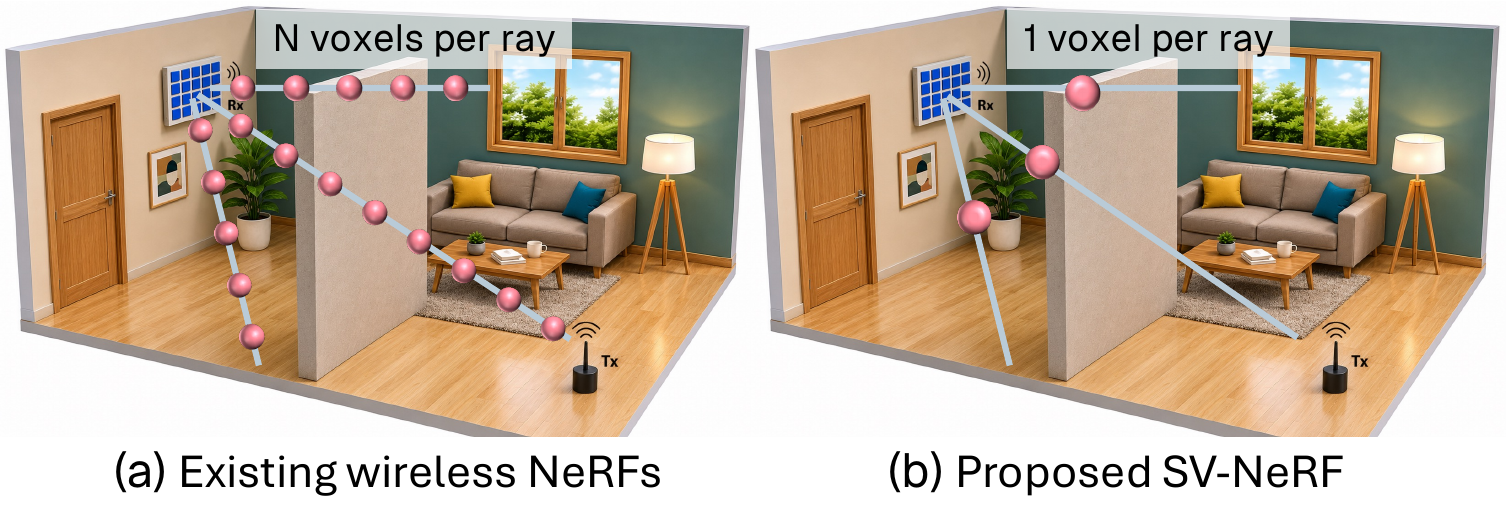}
    \caption{Existing wireless NeRFs rely on dense volumetric sampling along each propagation ray. The proposed single-voxel (SV) NeRF reduces rendering complexity by using a single voxel representation per ray.}
    \label{fig:intro_pic}
    \vspace{-6mm}
\end{figure}
Most existing wireless NeRF frameworks adopt volumetric rendering principles inherited from computer vision. Here, the 3D environment is discretized into multiple volumetric units, called \textit{voxels}, along each propagation \textit{ray}  (See Fig.~\ref{fig:intro_pic}(a)). 
Then, NeRF's neural networks estimate two parameters for each voxel: a \textit{volume density} and a \textit{radiance value}. The former represents the local occupancy (i.e., how much matter is present), whereas the latter models the view-dependent radiance emitted towards the RX. Together, these parameters implicitly capture volumetric RF propagation effects, including blockage-induced attenuation and multi-path components resulting from scattering and reflection within the environment. 
The resulting channel spatial spectrum is obtained by aggregating contributions from all voxels along the ray. This design implicitly assumes that dense voxelization is necessary to capture propagation phenomena and accurately reconstruct wireless channels. Consequently, wireless NeRF models often require large numbers of voxels and substantial computational resources for training the neural network pipeline and performing predictions when deployed~\cite{nerf2}.

Despite the widespread adoption of volumetric rendering, the role of voxel density in wireless NeRFs remains unexplored and not well-understood~\cite{nerf_wo_camera_neurips}. Unlike optical scenes, wireless propagation is dominated by a small number of physical paths whose contributions are ultimately observed through directional channel measurements~\cite{deepmimo}. This raises a fundamental question: 
\emph{Does wireless spatial spectrum synthesis  require dense volumetric representations, or are existing wireless NeRF models significantly over-parameterized?}



This paper presents the first systematic study of voxelization in wireless NeRF-based spatial spectrum synthesis. Specifically, we investigate how reducing the number of voxels per propagation ray affects spatial-spectrum magnitude prediction and model behavior. In vision NeRFs, learned density often provides a rough scene-structure signal, with larger values near visible surfaces or objects. Wireless NeRFs inherit this density-based representation, so one might expect the learned volume density to highlight RF-relevant structures, such as dominant paths, blockers, or reflectors. Instead, we find that the learned volume density has limited correspondence with dominant spatial-spectrum directions. This suggests that dense depth-wise voxel sampling may not provide the scene-level interpretability often assumed when NeRF-style rendering is adapted to wireless spatial spectrum magnitude prediction.

Motivated by this observation, we introduce two simplified wireless NeRF architectures, Single-Voxel NeRF$^2$ (SV-NeRF$^2$)
and Single-Voxel Instant Neural Graphics Primitives (SV-
INGP). These represent each propagation ray with a \emph{single-voxel} (SV), as we illustrate in  Fig.~\ref{fig:intro_pic}(b). We then evaluate the accuracy, rendering complexity, and training and inference times of the SV models relative to their multi-voxel counterparts.


The main contributions of this paper are as follows:

\begin{itemize}
\item We present the first systematic study of voxelization in wireless NeRF models and analyze whether it provides \textit{interpretable} information about learned volume density, number of voxels, and voxel placement along each spatial-spectrum direction.

\item We introduce SV-NeRF$^2$ and SV-INGP as \emph{single-voxel} variants of NeRF$^2$~\cite{nerf2} and INGP~\cite{ingp} for wireless spatial spectrum magnitude prediction. To the best of our knowledge, this is the first work to investigate both multi-voxel and single-voxel INGP variants in the wireless domain.

\item  On the real-world NeRF$^2$ dataset, SV-INGP matches NeRF$^2$'s median SSIM of $0.78$ with 184x lower training time. Isolating the sampling change within INGP, SV-INGP reduces training and inference time by 8x and 10x, respectively, with only a 2.5$\%$ SSIM reduction compared to multi-voxel INGP.

\end{itemize}

\section{Related Work}

\textbf{NeRF models} were first introduced for novel view synthesis in computer vision by learning continuous density and radiance fields from images~\cite{nerf,mip_nerf}. This idea has recently been adapted to wireless propagation modeling, where the goal is to synthesize RF spatial spectrum magnitude from sparse measurements. NeRF$^2$ introduced one of the first wireless NeRF frameworks by combining volumetric rendering with wireless propagation models~\cite{nerf2}. Follow-up works have explored wireless scene reconstruction, sensing, and RF digital twins using NeRF-style representations~\cite{newrf,nerf-apt,NeRF2-ss,rfcanvas}. Related neural field ideas have also appeared in other domains, such as radar imaging~\cite{dart_radar_nerf}.

A common design choice across these works is dense volumetric rendering: each propagation ray is sampled using many voxels, and the predicted channel spatial spectrum is obtained by aggregating their contributions. This design is natural for vision, where dense scene structure is central to image formation. In wireless, however, propagation is often dominated by a few strong paths, and it is unclear whether dense depth-wise voxelization is needed for spatial spectrum magnitude prediction. Our work directly studies this question and shows that a single-voxel representation can preserve prediction accuracy while reducing rendering cost.

\textbf{Vision-based INGP} introduced multi-resolution hash encoding to accelerate NeRF training while preserving high reconstruction quality~\cite{ingp}. We build on this efficient encoding for wireless scenes, but our goal is different. INGP reduces the cost of neural representation learning while retaining dense ray sampling. In contrast, we ask whether dense sampling itself is necessary, providing an orthogonal path to reducing rendering complexity.

\textbf{Gaussian Splatting} represents scenes using learnable Gaussian primitives rather than volumetric density fields~\cite{gaussian_splat_cv}. Recent wireless systems have adopted this idea for RF channel spatial spectrum prediction and scene reconstruction~\cite{wrf_gs_wen_2025,gsparc_tamu,gs_rf-3dgs_60GHz}. Although promising for fast rendering, Gaussian Splatting represents a fundamentally different modeling philosophy than NeRF-based approaches. NeRFs explicitly learn density and radiance functions that provide a volumetric representation of wireless propagation. This structure provides a natural framework for studying how scene representations contribute to spatial-spectrum synthesis. Our work, therefore, focuses on NeRF-based models and investigates the role of volumetric density itself.

\section{Single-voxel Wireless NeRF Design}
\label{sec:singe-voxel-wireless-NeRF}
In this section, we present a single-voxel wireless NeRF model that serves a twofold purpose. First, it is a novel approach to the NeRF method that achieves performance comparable to existing related methods while incurring significantly lower computational cost. Second, it provides context for assessing the value of voxel density. 

\subsection{RF Scene Model}
\label{subsec:rf_scene_model}

\begin{figure} [!htbp]
\centering
\includegraphics[width=\linewidth,
trim={2.0in 2.3in 2.5in 2.6in},
clip]{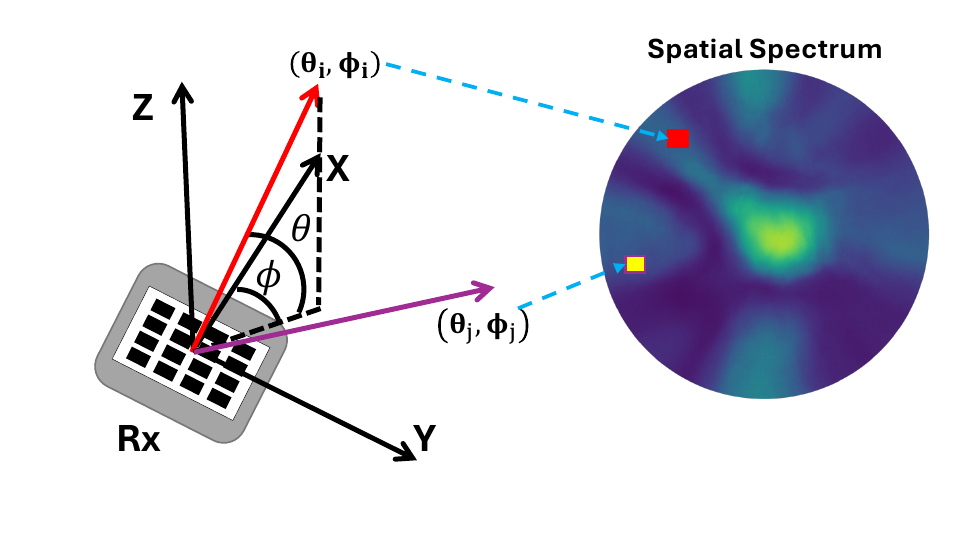}
\caption{RX coordinate convention and spatial spectrum in polar form.}
\label{fig:system-model}
\vspace{-4mm}
\end{figure}

Consider a static scene with a fixed $K \times K$ planar RX array and a single-antenna TX placed at different locations. For each TX position, the RX records a $K \times K$ complex channel response. Fig.~\ref{fig:system-model} describes the angular domain convention at the receiver side, where each candidate arrival direction is denoted by $\omega=(\theta,\phi)$, where $\theta$ and $\phi$ are the elevation and azimuth angles respectively, with unit vector
$\mathbf{u}(\omega)
=
[\cos\theta\cos\phi\; ,\; \cos\theta\sin\phi,\;\sin\theta]^\top$.

We sample the front-facing angular domain using $\theta\in[0^\circ,90^\circ]$ and $\phi\in[0^\circ,360^\circ]$ with $1^\circ$ resolution in both dimensions. To obtain a direction-dependent representation, we project the array-domain channel response onto steering vectors over this angular grid. For each direction $\omega$, the spatial spectrum is computed as
\begin{align}
\psi(\omega)=
\Big|
\frac{1}{K^2}
\sum_{n=1}^{K^2}
H_n
\exp\left(-j2\pi \; \mathbf{r}_n^\top \cdot \mathbf{u}(\omega)\right)
\Big|\  ,
\label{eq:spatial-spectrum}
\end{align}
where $H_n$ is the complex channel at the $n$-th RX antenna, $\mathbf{r}_n$ is a vector that represents the corresponding antenna position with respect to an antenna origin normalized by wavelength and $\mathbf{u}(\omega)$ is the unit vector associated with direction $\omega$. The resulting spectrum $\psi(\omega)$ represents the magnitude of the beamformed channel response as a function of direction, with dominant peaks corresponding to directions from which strong signal components arrive. Each angular direction in the spatial spectrum defines a ray originating at the receiver and extending into the scene. For a direction $\omega$, the corresponding ray is parameterized as
$\mathbf{x}(l,\omega)=P_{\mathrm{RX}}+l\mathbf{u}(\omega)$, where $l\in[0,d]$,
$P_{\mathrm{RX}}$ is the RX location, and $d$ is the maximum ray depth. Existing NeRF models sample multiple points along this ray and accumulate their contributions to estimate the channel response along direction $\omega$. For a sample at depth $l$, the explicit free-space propagation factor from the sample to the receiver is modeled as
\begin{align}
\mathrm{PL}(l)
=
\frac{c}{4\pi f_c l}
\exp\left(-j\frac{2\pi f_c}{c}l\right),
\label{eq:free-space-path-loss}
\end{align}
where $c$ is the speed of light and $f_c$ is the carrier frequency. These definitions provide the notation used in the next section to formulate the proposed single-voxel NeRF model.

For each TX location, a $90 \times 360$ spatial spectrum image is generated
using~\eqref{eq:spatial-spectrum}. Each spectrum is then normalized by its own
maximum pixel value,
\begin{align}
\tilde{\psi}(\omega)
=
\frac{\psi(\omega)}{\max_{\omega'} \psi(\omega')} \ ,
\end{align}
rather than by a single global scale factor shared across the dataset. In the remainder of the paper, $\psi(\omega)$ refers to this normalized spectrum, which acts as the ground truth for the network. The spectra are linear amplitudes, and amplitude decays as $1/l$ with propagation
distance, as in~\eqref{eq:free-space-path-loss}. Under a single global scale,
spectra from TX locations near the RX dominate the range and keep their angular
structure, while spectra from distant TX locations are compressed close to zero
and contribute little to training. Per-spectrum normalization removes this level difference, so that spectra from all TX locations contribute comparably during training.

\subsection{Proposed Single-Voxel INGP}
\begin{figure} [!htbp]
\centering
\includegraphics[width=\linewidth,
trim={1.0in 3.3in 0.8in 0.5in},
clip]{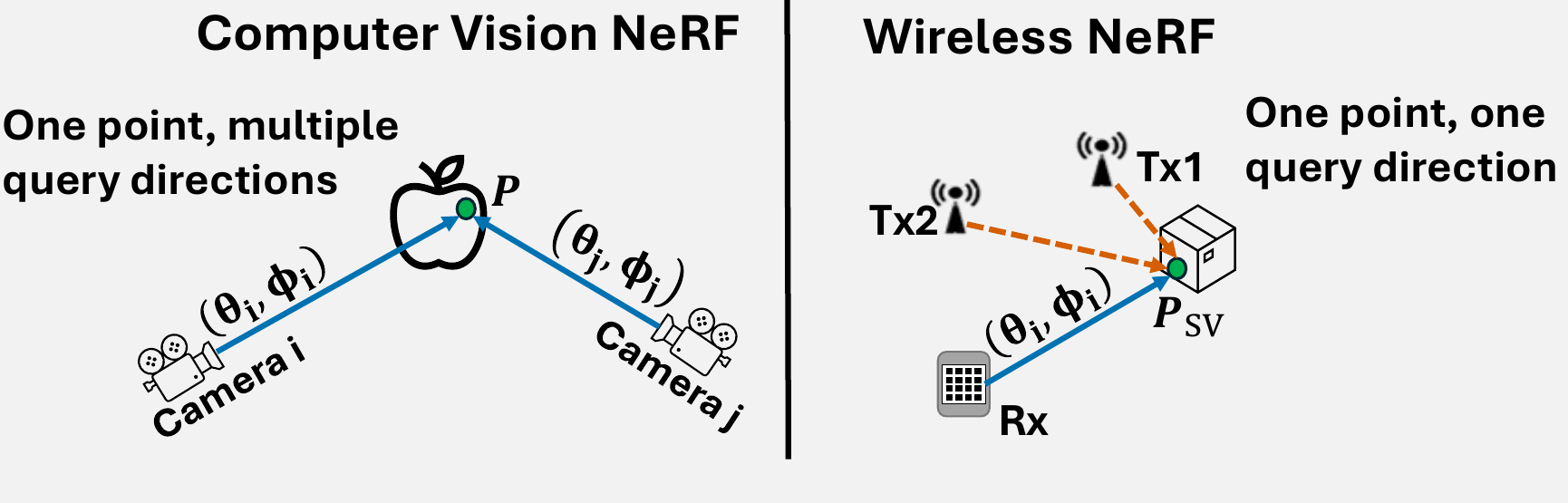}
\caption{In computer vision, a location in the scene is sampled by different camera poses. For wireless, a location in the scene is sampled by the fixed multi-antenna Rx position, which defines a fixed set of ray directions independent of the Tx locations.}
\label{fig:cv-vs-wireless-comparison}
\vspace{-4mm}
\end{figure}

\textbf{Motivation.} Existing RF NeRF-based models evaluate several locations along each ray and accumulate their contributions to generate the spatial spectra, similar to vision NeRFs~\cite{nerf2, newrf}. Instead, we evaluate a single location along each ray and show that it achieves similar high performance. A natural question is how a single location per ray is enough to model wireless environments. Our intuition is motivated by differences in measurement geometry between the vision and wireless settings. In a wireless setting, the RX array is stationary, so every ray originates from the same Rx location $P_{\rm RX}$, and any sampled location $\mathbf{x}(l,\omega)$ is determined entirely by its depth $l$ and the direction $\omega$. Each voxel in the scene, therefore, lies on exactly one specific ray direction and contributes to exactly one pixel of the spectrum. In vision NeRFs, the same point is crossed by rays from many camera poses, and each of those views constrains the quantities assigned to it. Fig.~\ref{fig:cv-vs-wireless-comparison} illustrates the difference between a sampled location in the computer vision and wireless domains. That shared constraint is what justifies the use of the multi-voxel depth sampling along each direction. In contrast, in our wireless setting, a fixed receiver provides no such constraint. A voxel in a particular direction is not queried by any other direction, regardless of how many voxels we place along that ray direction.  The measurement along direction $\omega$ constrains only the accumulated contribution of the ray, and any distribution of that contribution over depth, either with one voxel per ray or multiple voxels per ray, is consistent with it. We therefore place a single voxel on each ray and let the network represent the aggregate.

\begin{figure*}[!t]
    \centering
    \includegraphics[width=\textwidth,
    trim={0.0in 3.1in 0.0in 0.0in},
        clip]
    {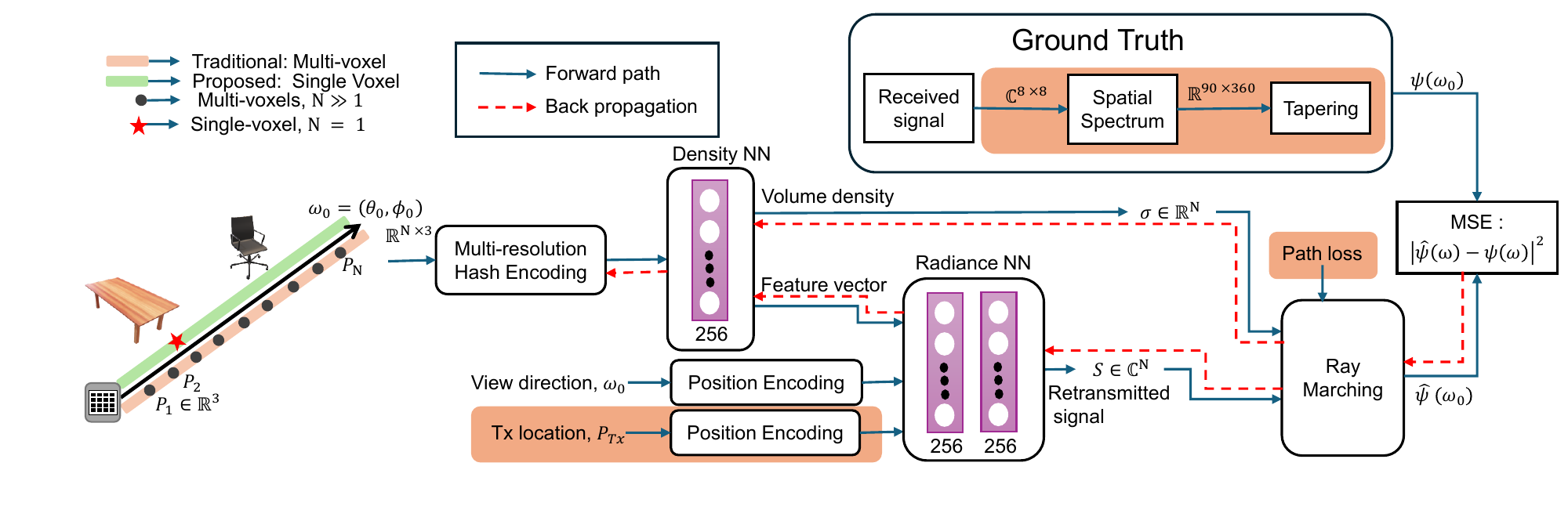}
    \caption{Overview of SV-INGP. The left block compares multi-voxel INGP with the proposed single-voxel sampling, while the right block shows the training pipeline for spatial-spectrum magnitude prediction. \colorbox{orange!60}{Orange} blocks highlight wireless-specific modifications relative to vision-based INGP~\cite{ingp}.}
    \label{fig:ingp-wireless-block-diagram}
    \vspace{-6mm}
\end{figure*}

\textbf{Design.} Next, we present the design of using a single voxel approach in the context of INGP-based models. Nevertheless, the same principle applies to other models such as NeRF$^2$, which we will not discuss for brevity. The block diagram of our proposed approach, SV-INGP, is shown in Fig.~\ref{fig:ingp-wireless-block-diagram}. The highlighted blocks show the changes made to the vision-INGP to adapt to our wireless setting. 
The SV-INGP pipeline is designed as follows: it takes as input the transmitter location $P_{\rm TX}$, a spatial direction $\omega$, and the corresponding single-voxel location $P_{\rm sv}$ on the ray associated with $\omega$ and predicts the spatial-spectrum magnitude $\hat{\psi}(\omega)$, which is trained against the ground-truth magnitude $\psi(\omega)$. The RX location and orientation, and scene bounds are fixed for each experiment and are used only to define the coordinate system, ray geometry, and voxel placement. These quantities are provided at initialization and are not treated as network inputs. This framework is enabled by two multi-layer perceptrons (MLPs), one for volume density prediction and another for radiance field prediction (complex retransmitted signal). Using the ray parameterization from the previous section, each spatial direction $\omega$ defines a ray $\mathbf{x}(l,\omega)$ originating from the receiver. SV-INGP evaluates only one point on this ray, denoted by
$P_{\rm sv}=\mathbf{x}(l_{\rm sv},\omega)$,
where $l_{\rm sv}$ is the depth of the selected voxel. The voxel location $P_{\rm sv}$ is passed to the density MLP, which outputs a scalar volume density $\sigma(P_{\rm sv})$ and a feature vector. The feature vector, together with the direction $\omega$ and transmitter location $P_{\rm TX}$, is then passed to the radiance MLP, which outputs the complex retransmitted signal $S(P_{\rm sv},\omega,P_{\rm TX})$. The learned voxel contribution is combined with the explicit propagation factor $\mathrm{PL}(l_{\rm sv})$ to form the following estimate:
\begin{equation} 
\hat{\psi}(\omega) =
\left|
\alpha\left(\sigma(P_{\rm sv})\right)
\cdot S(P_{\rm sv}, \omega, P_{\rm TX})
\cdot {\rm PL}(l_{\rm sv})
\right| .
\label{eq:single-voxel-ingp-power}
\end{equation}
Since SV-INGP assigns a single voxel to represent the entire ray segment, the effective voxel thickness is set to the maximum ray depth, i.e., $\Delta_{\rm sv}=d$. The attenuation weight in~\eqref{eq:single-voxel-ingp-power} is then defined as
\begin{align}
\alpha(\sigma(P_{\rm sv}))
=
1-\exp\left(-\sigma(P_{\rm sv})\Delta_{\rm sv}\right)\ ,
\end{align}
where $\sigma(P_{\rm sv})\in[0,\infty)$ is the predicted volume density at the single-voxel location. This maps the density value to a bounded weight $\alpha(\sigma(P_{\rm sv}))\in[0,1)$. The model is trained to match the ground-truth spatial-spectrum magnitude $\psi(\omega)$. For a batch of $M$ directions, we use the \textbf{\textit{mean-squared error loss}}
\begin{equation}
\mathcal{L}_{\rm MSE}
=
\frac{1}{M}
\sum_{i=1}^{M}
\left(
\hat{\psi}(\omega_i)-\psi(\omega_i)
\right)^2\ .
\label{eq:loss-function}
\end{equation}

\textbf{Encoding.} Before the inputs are passed to the MLPs, they are mapped to higher-dimensional representations. For the direction $\omega$ and transmitter location $P_{\rm TX}$, we use the sinusoidal positional encoding used in NeRF~\cite{nerf}. For a scalar input component $q$, the encoding is
\begin{equation}
    \begin{aligned}
    \gamma(q) =&
    [\sin(2^0\pi q), \sin(2^1\pi q),\ldots, \sin(2^{L-1}\pi q),\\
    & \cos(2^0\pi q), \cos(2^1\pi q),\ldots ,\cos(2^{L-1}\pi q)] ,
    \end{aligned}
\end{equation}
where $L$ is the number of frequency bands. Thus, a scalar input is mapped to $\mathbb{R}^{2L}$, and a 3D input is mapped to $\mathbb{R}^{6L}$ when the encoding is applied independently to each coordinate and concatenated.

For the single-voxel location $P_{\rm sv}$, we use the multi-resolution hash encoding introduced in INGP~\cite{ingp}. While INGP was originally designed for dense multi-sample ray rendering, we adopt a similar idea in the single-voxel setting. Our intuition is to define a global multi-resolution grid over the scene that is independent of the location of each voxel on each ray. The selected voxel location $P_{\rm sv}$ is then encoded by querying this global grid across multiple coarse and finer resolutions. This design preserves the compact high-resolution scene representation provided by INGP, while allowing SV-INGP to evaluate only one learned spatial sample per ray.

\subsection{Multi-Voxel INGP Baseline}

To assess the value of voxel density, we also create a multi-voxel counterpart to SV-INGP. Such an extension requires placing multiple samples per ray and accumulating their contributions. 
 
For $N$ voxels per ray, the single-voxel estimate in~\eqref{eq:single-voxel-ingp-power} becomes
\begin{equation}
\hat{\psi}(\omega) =
\Big|
\sum_{j=1}^N
\alpha\left(\sigma(P_j)\right)
\cdot S(P_j, \omega, P_{\rm TX})
\cdot {\rm PL}(l_j)
\Big|\ ,
\label{eq:multi-voxel-ingp-power}
\end{equation}
where $P_j=\mathbf{x}(l_j,\omega)$ is the location of the $j$-th voxel along the ray and $l_j$ is its distance from the receiver. 

\subsection{Insights into single-voxel model}
\label{subsec:sv-insights}
Under the fixed-receiver geometry, the multi-voxel model
in~\eqref{eq:multi-voxel-ingp-power} is a reparameterization of the same set of predictions as provided by~\eqref{eq:single-voxel-ingp-power} rather than a more expressive one.

\textbf{Why single-voxel per ray works?} Let $C_{\rm MV}(\omega,P_{\rm TX})$ and $C_{\rm SV}(\omega,P_{\rm TX})$ denote
the complex sums inside the magnitude
in~\eqref{eq:multi-voxel-ingp-power} and~\eqref{eq:single-voxel-ingp-power},
respectively. The voxel depth $l_{\rm sv}$ is fixed at initialization, so
$P_{\rm sv}=\mathbf{x}(l_{\rm sv},\omega)$ is determined by $\omega$, and the
density MLP takes only $P_{\rm sv}$ as input. Both
$\alpha(\sigma(P_{\rm sv}))$ and ${\rm PL}(l_{\rm sv})$ are therefore functions
of $\omega$ alone. In particular, neither depends on $P_{\rm TX}$: together they
form a fixed per-direction factor that is identical for every transmitter
location. The radiance network, by contrast, receives $\omega$ and
$P_{\rm TX}$ directly, so its output is free to take any value as a function of
$(\omega,P_{\rm TX})$. Setting 
\begin{equation}
S(P_{\rm sv},\omega,P_{\rm TX})
=
\frac{C_{\rm MV}(\omega,P_{\rm TX})}
{\alpha(\sigma(P_{\rm sv}))\,{\rm PL}(l_{\rm sv})} ,
\end{equation}
gives $C_{\rm SV}=C_{\rm MV}$ for every direction and transmitter location whenever $\alpha(\sigma(P_{\rm sv}))>0$. The single-voxel model can therefore represent any prediction the multi-voxel model can. The single-voxel model can be thought of as a special case of multi-voxel model, where the signal contribution (either volume density or radiance output) of all but one voxel is set to zero. So the two models are equivalent in what they can express, up to the approximation capacity of the network.

\textbf{Impact of normalization:} The normalization choice in Section~\ref{subsec:rf_scene_model} further
favors the single-voxel design. Under a global scale, the target carries the absolute signal level, which decays with transmitter distance, and the multi-voxel model can produce this decay by shifting weight across voxels whose path-loss factors ${\rm PL}(l_j)$ already span a range of depth-dependent gains. The single-voxel model has one fixed ${\rm PL}(l_{\rm sv})$ per direction that does not vary with $P_{\rm TX}$, so its radiance network must learn the same decay on its own. Normalizing each spectrum by its own maximum removes absolute level from the target and leaves only the angular distribution of energy. What remains is a per-direction quantity that a single voxel represents as directly as a full depth sweep.

\section{Evaluations}
\label{sec:empirical evaluations}

\begin{figure*}[!t]
    \centering
    \footnotesize
    \setlength{\tabcolsep}{2pt}
    \renewcommand{\arraystretch}{1.0}

    \begin{minipage}[t]{0.25\textwidth}
        \vspace{0pt}
        \centering


        \begin{subfigure}[t]{1\linewidth}
            \centering
            \includegraphics[width=\linewidth]{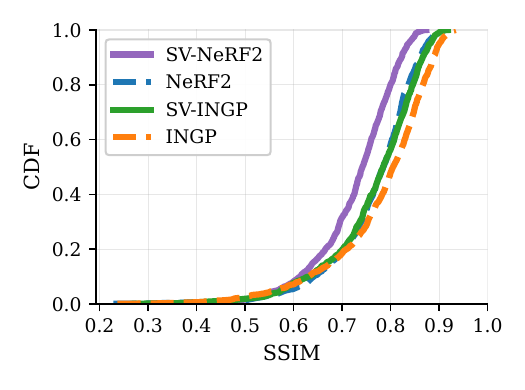}
            \caption{SSIM comparison}
            \label{fig:realworld-ssim-cdf}
        \end{subfigure}

        \begin{subfigure}[t]{.85\linewidth}
            \centering
            \includegraphics[width=\linewidth]{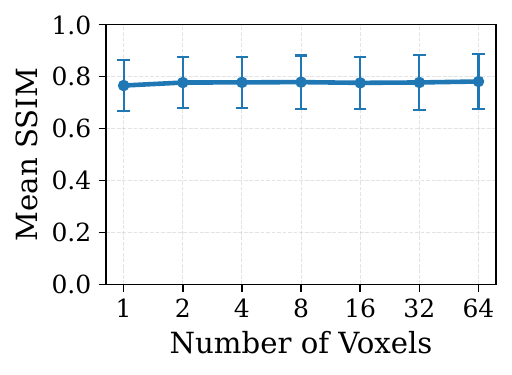}
            \caption{Impact of no. of Voxels}
            \label{fig:realworld-scene}
        \end{subfigure}
    \end{minipage}
    \hfill
    \begin{subfigure}[t]{0.73\textwidth}
        \vspace{0pt}
        \centering
        \begin{tabular}{>{\bfseries}c | c | c c c c}
            \toprule
            \textbf{TX Positions}
            & \textbf{Ground Truth}
            & \textbf{SV-INGP}
            & \textbf{INGP}
            & \textbf{SV-NeRF$^2$}
            & \textbf{NeRF$^2$} \\
            \midrule

            \raisebox{0.045\textwidth}{P1} &
            \includegraphics[width=0.105\textwidth]{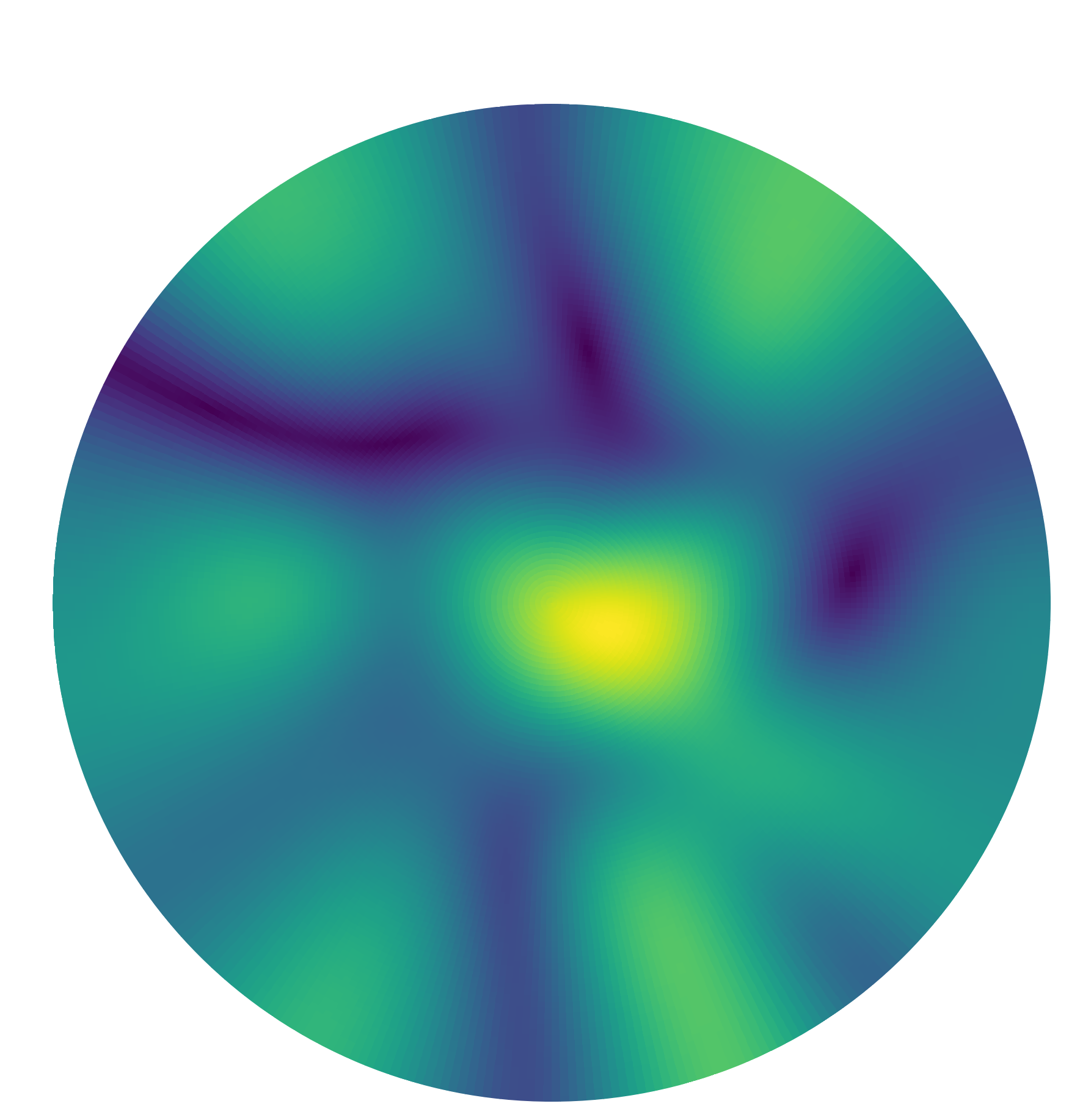} &
            \includegraphics[width=0.105\textwidth]{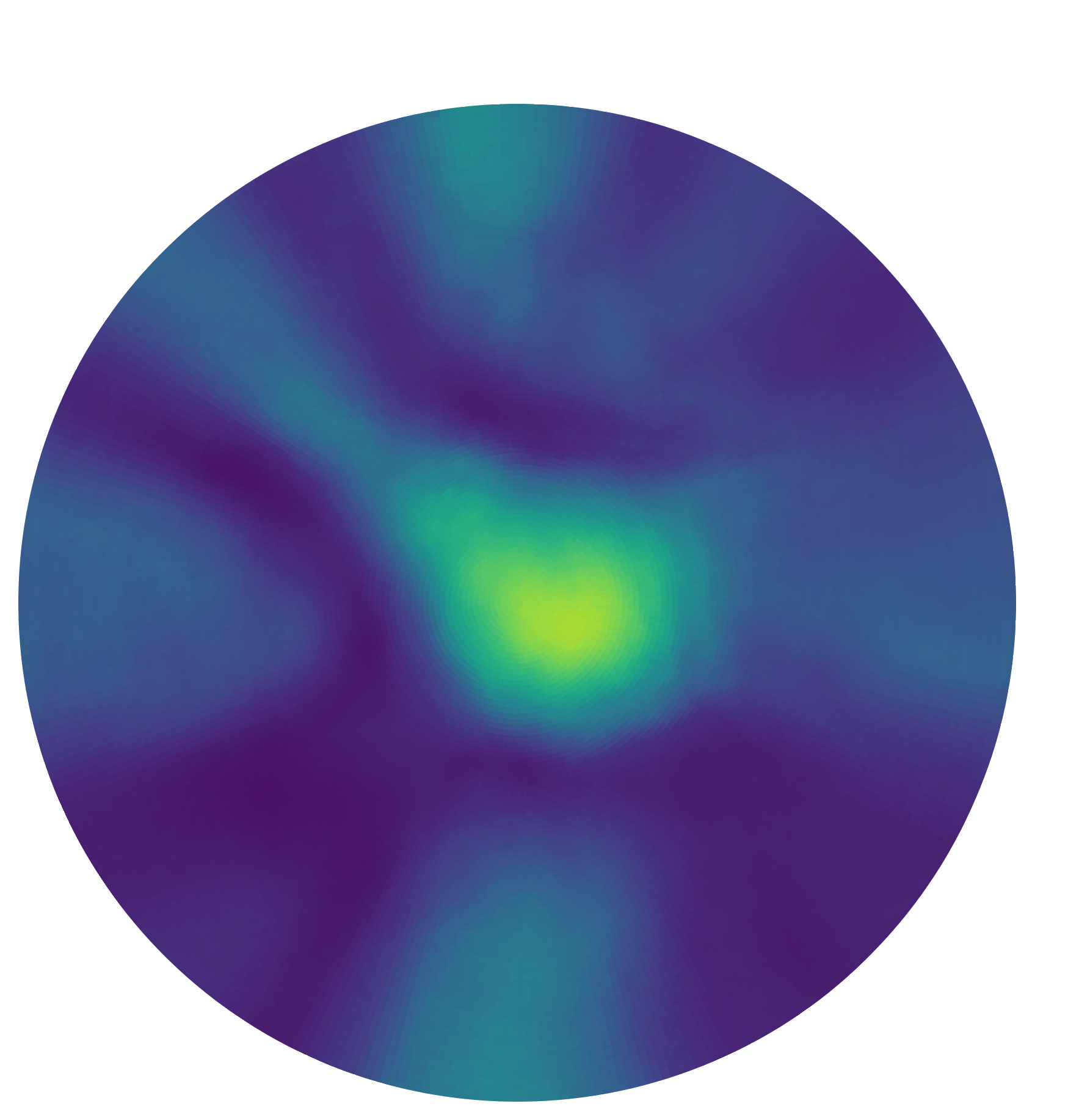} &
            \includegraphics[width=0.105\textwidth]{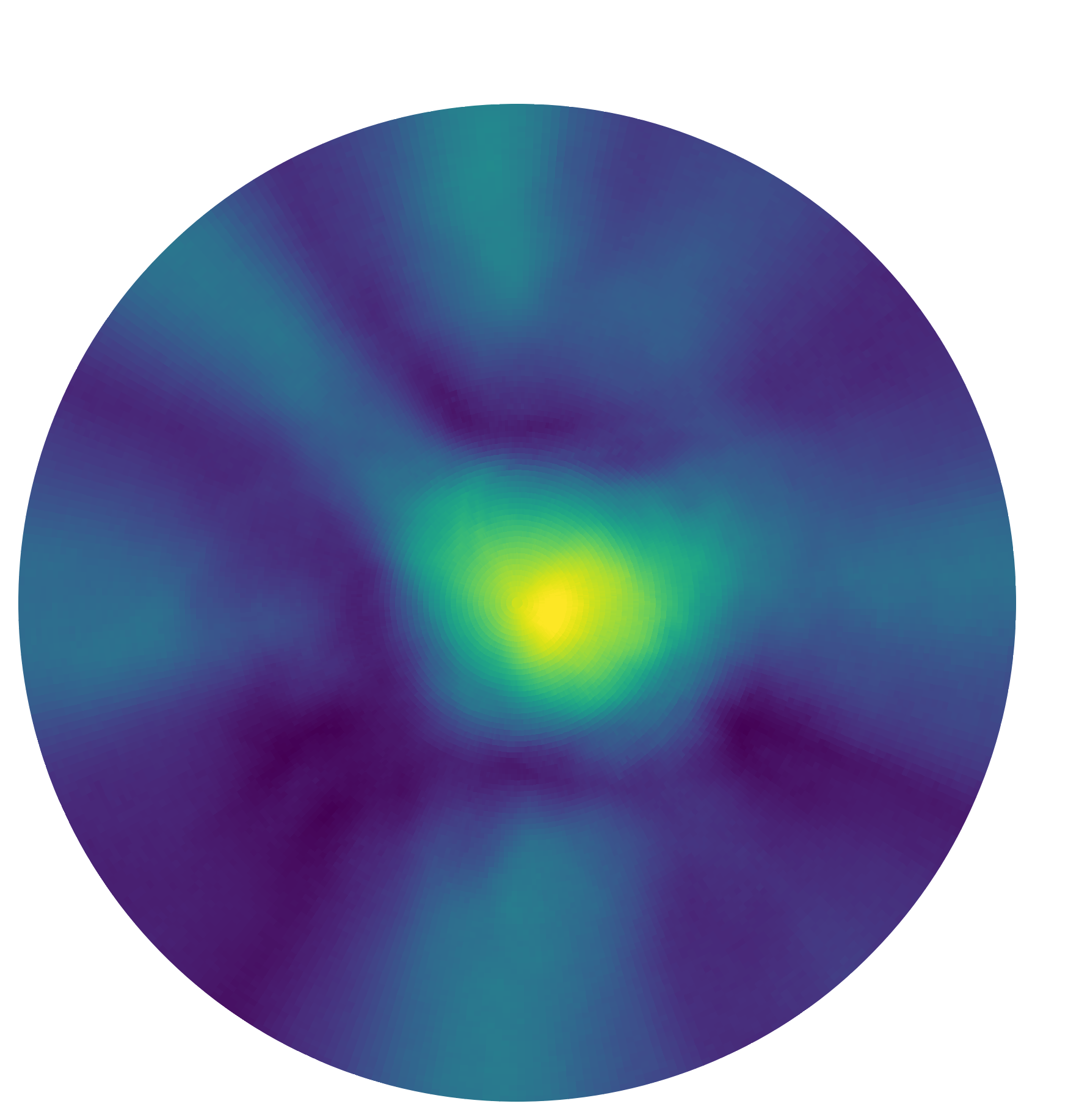} &
            \includegraphics[width=0.105\textwidth]{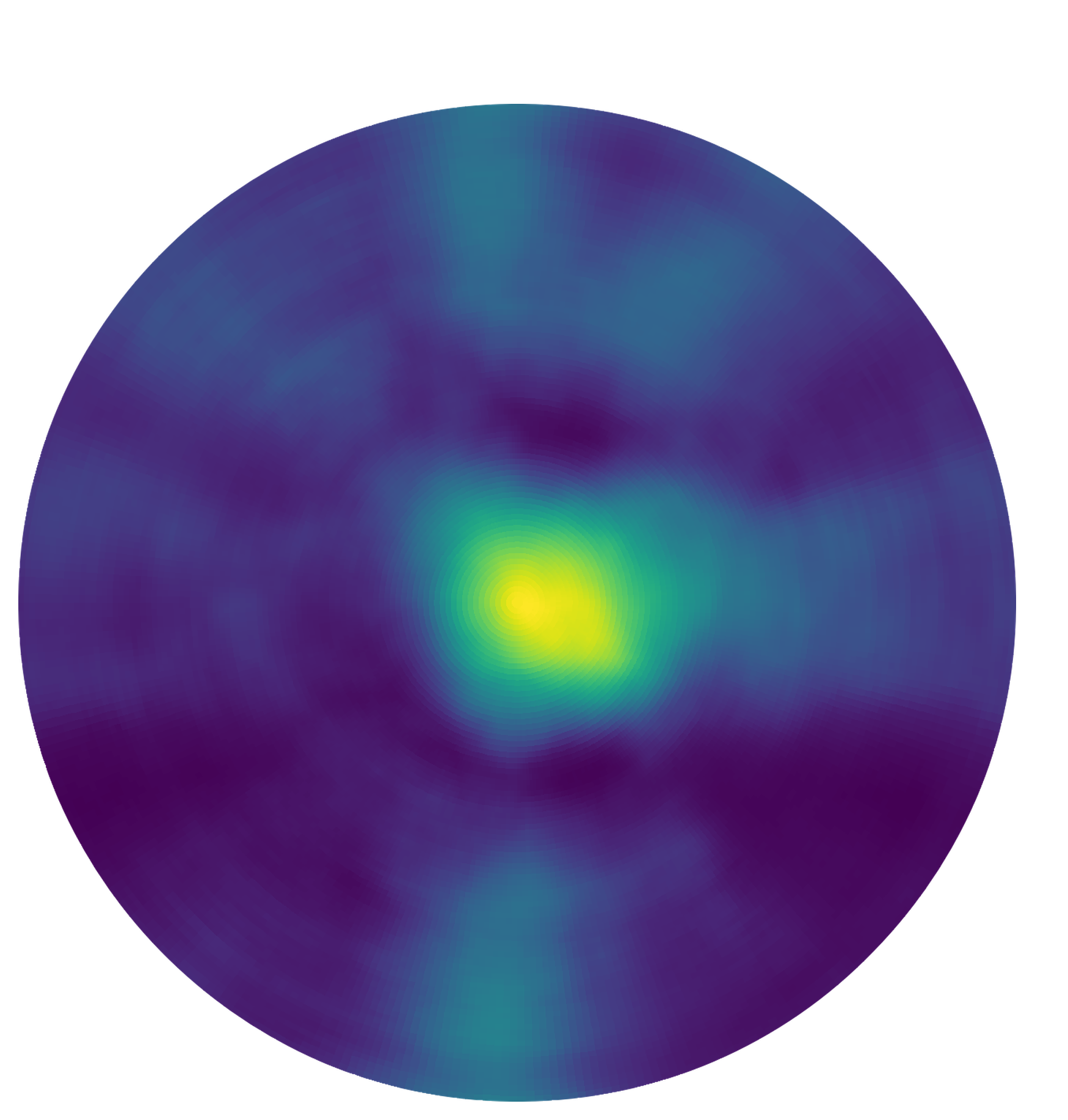} &
            \includegraphics[width=0.105\textwidth]{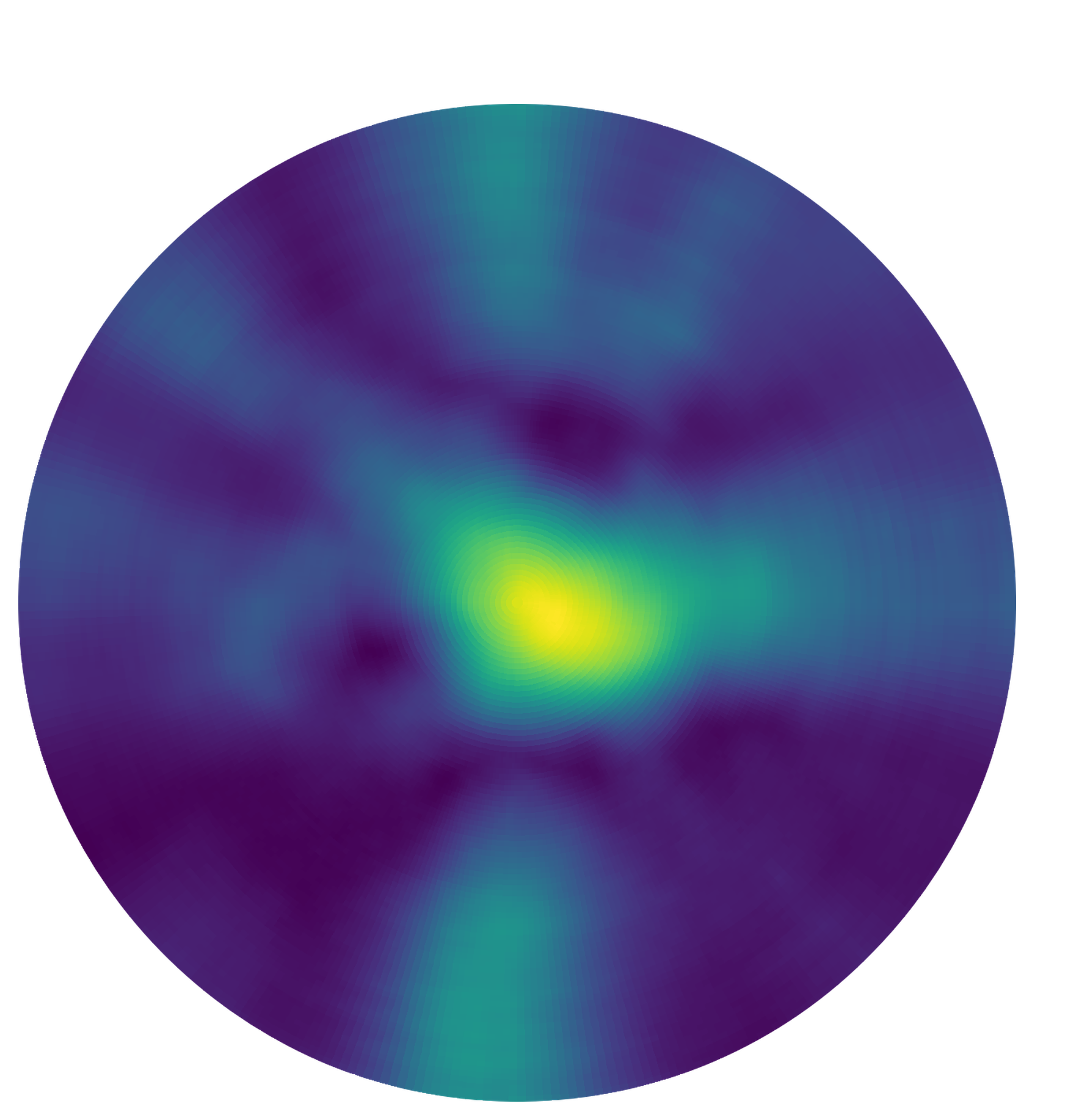} \\

            \raisebox{0.045\textwidth}{P2} &
            \includegraphics[width=0.105\textwidth]{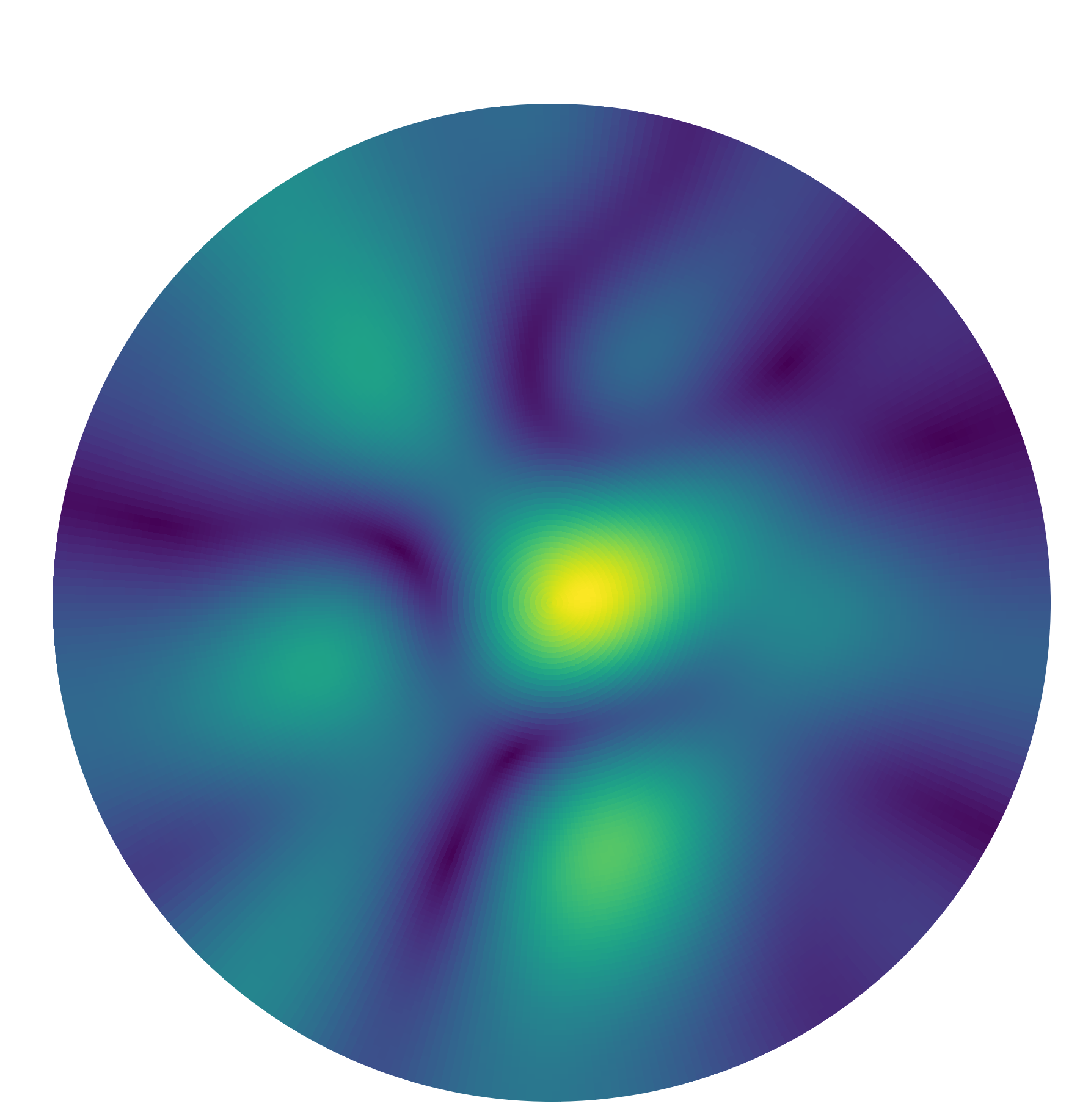} &
            \includegraphics[width=0.105\textwidth]{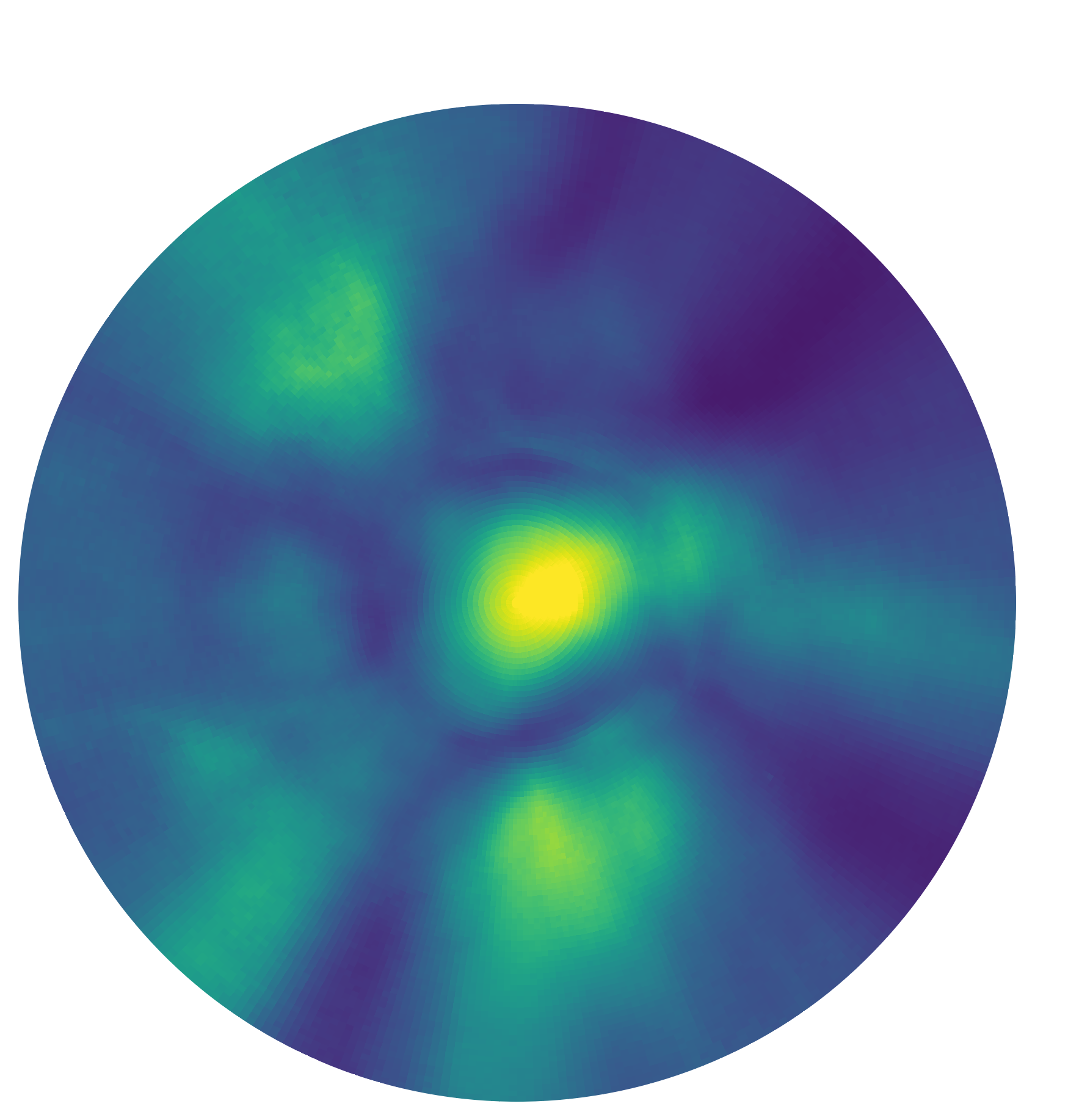} &
            \includegraphics[width=0.105\textwidth]{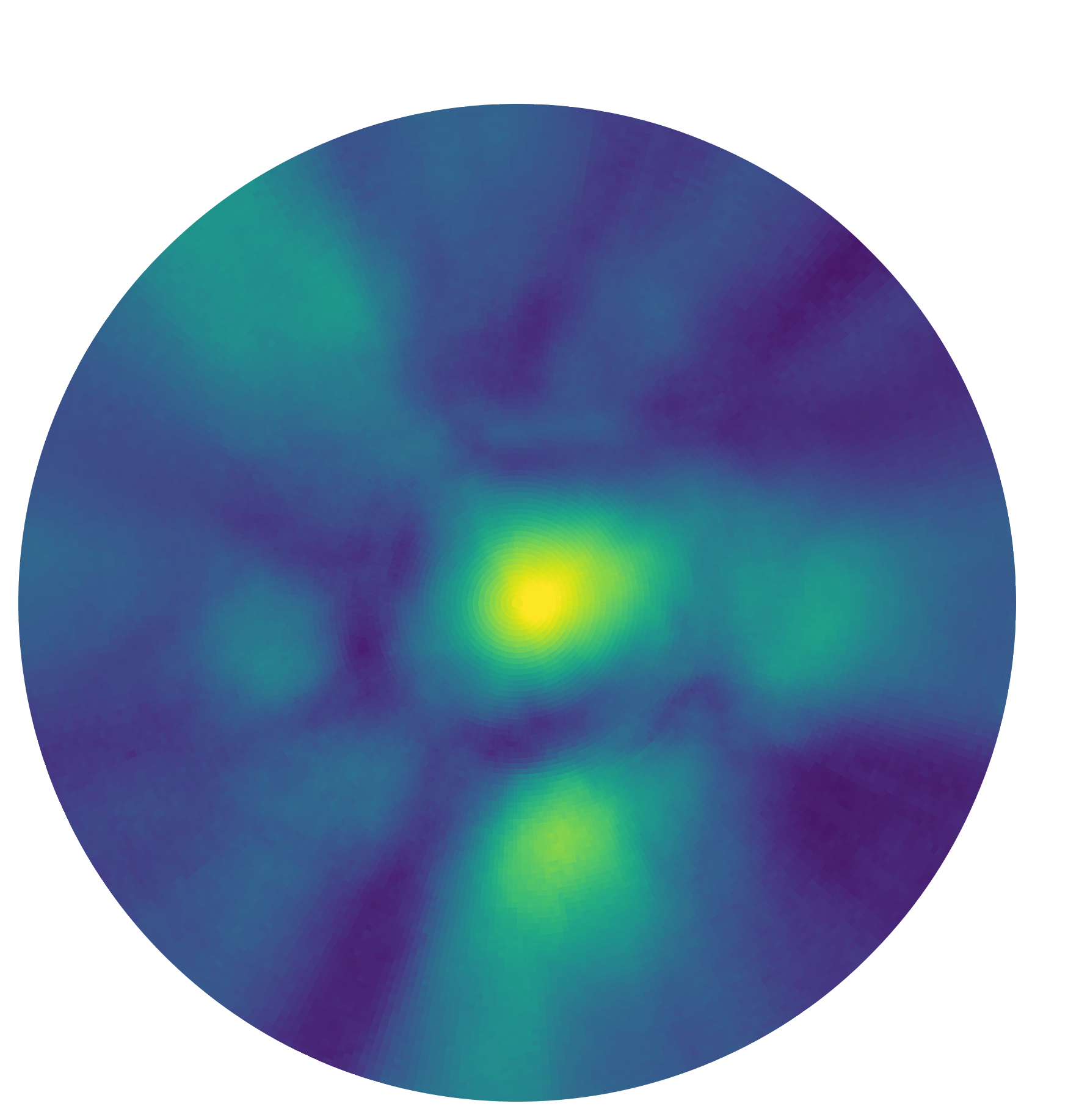} &
            \includegraphics[width=0.105\textwidth]{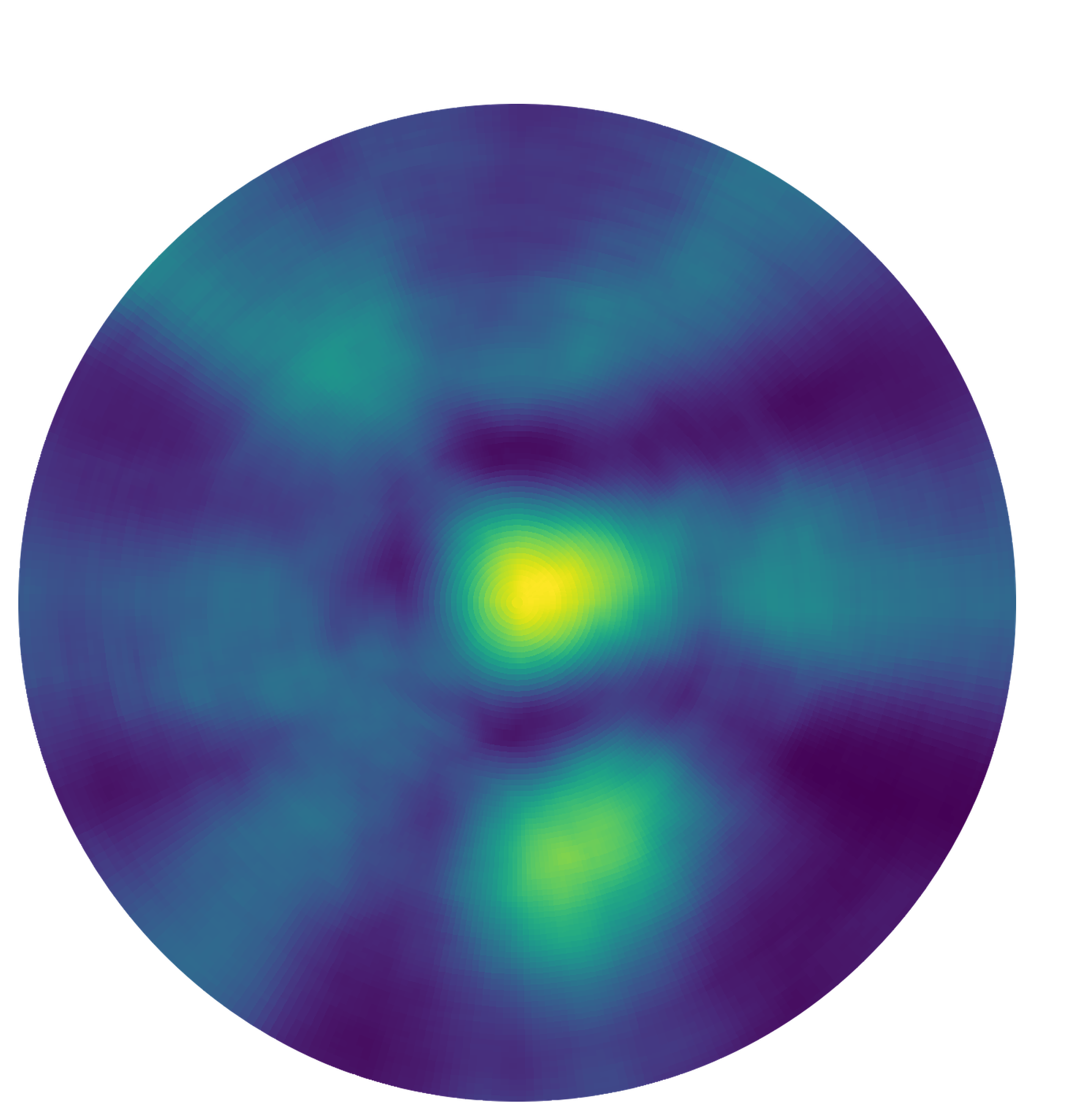} &
            \includegraphics[width=0.105\textwidth]{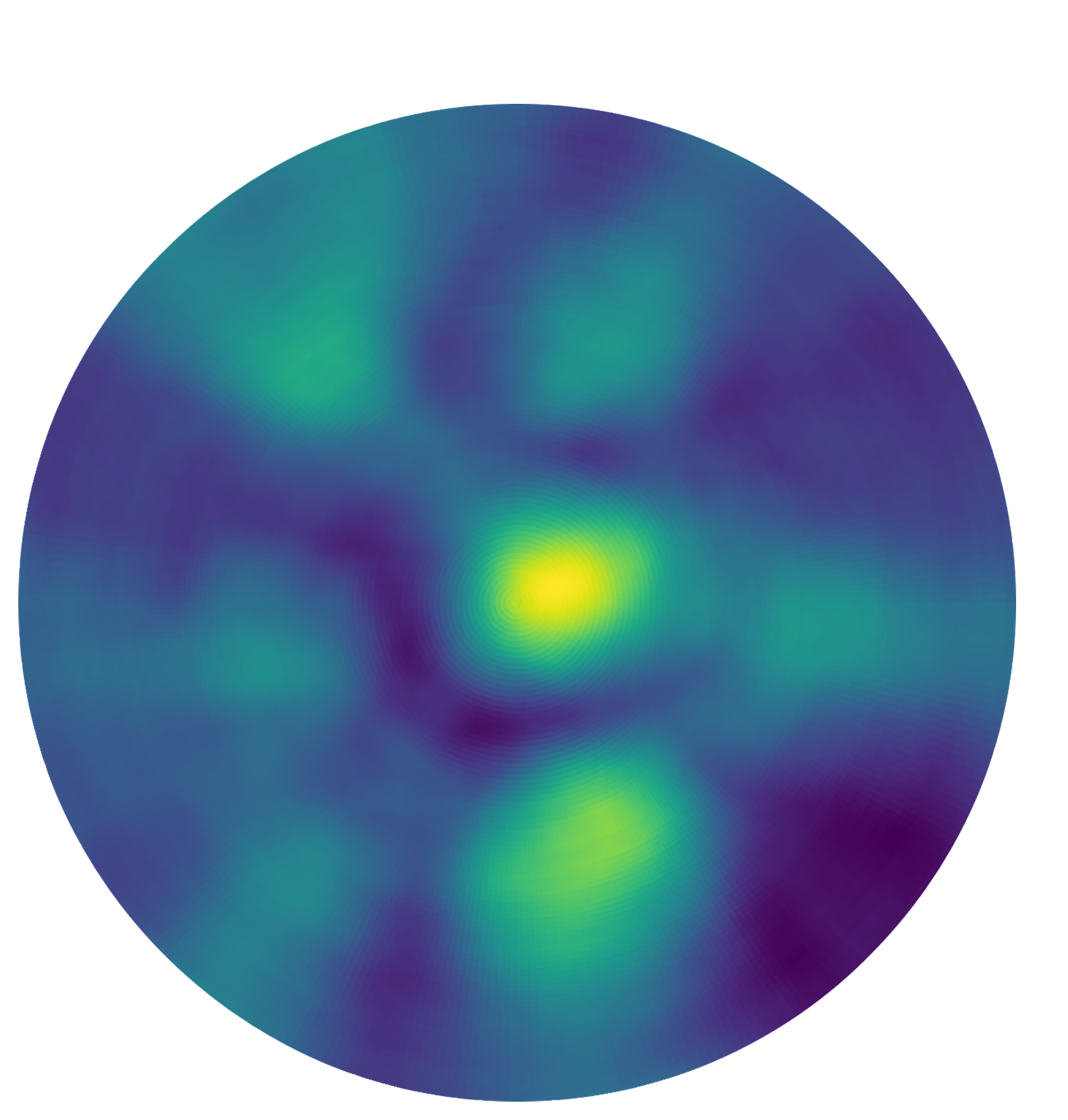} \\

            \raisebox{0.045\textwidth}{P3} &
            \includegraphics[width=0.105\textwidth]{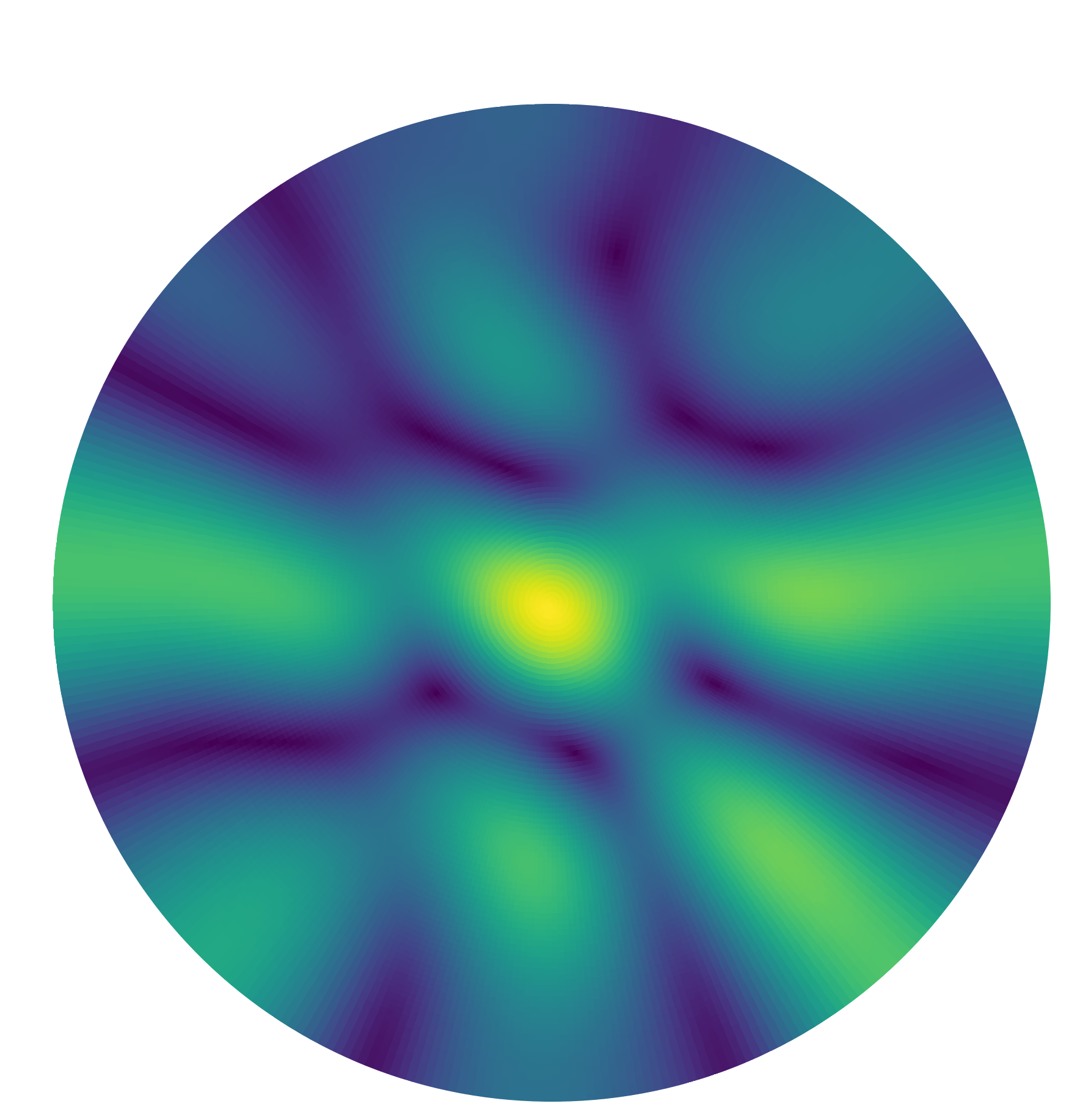} &
            \includegraphics[width=0.105\textwidth]{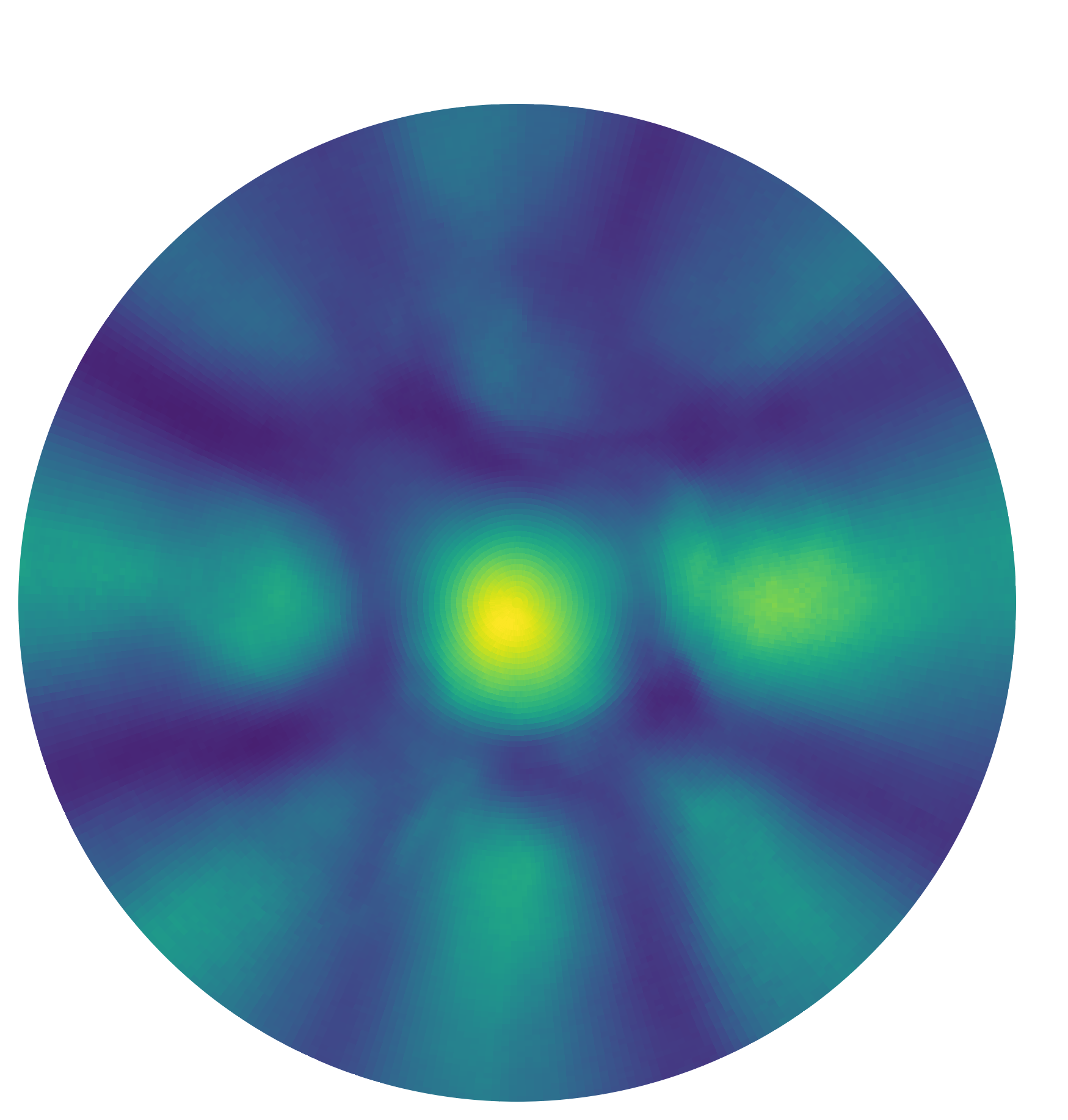} &
            \includegraphics[width=0.105\textwidth]{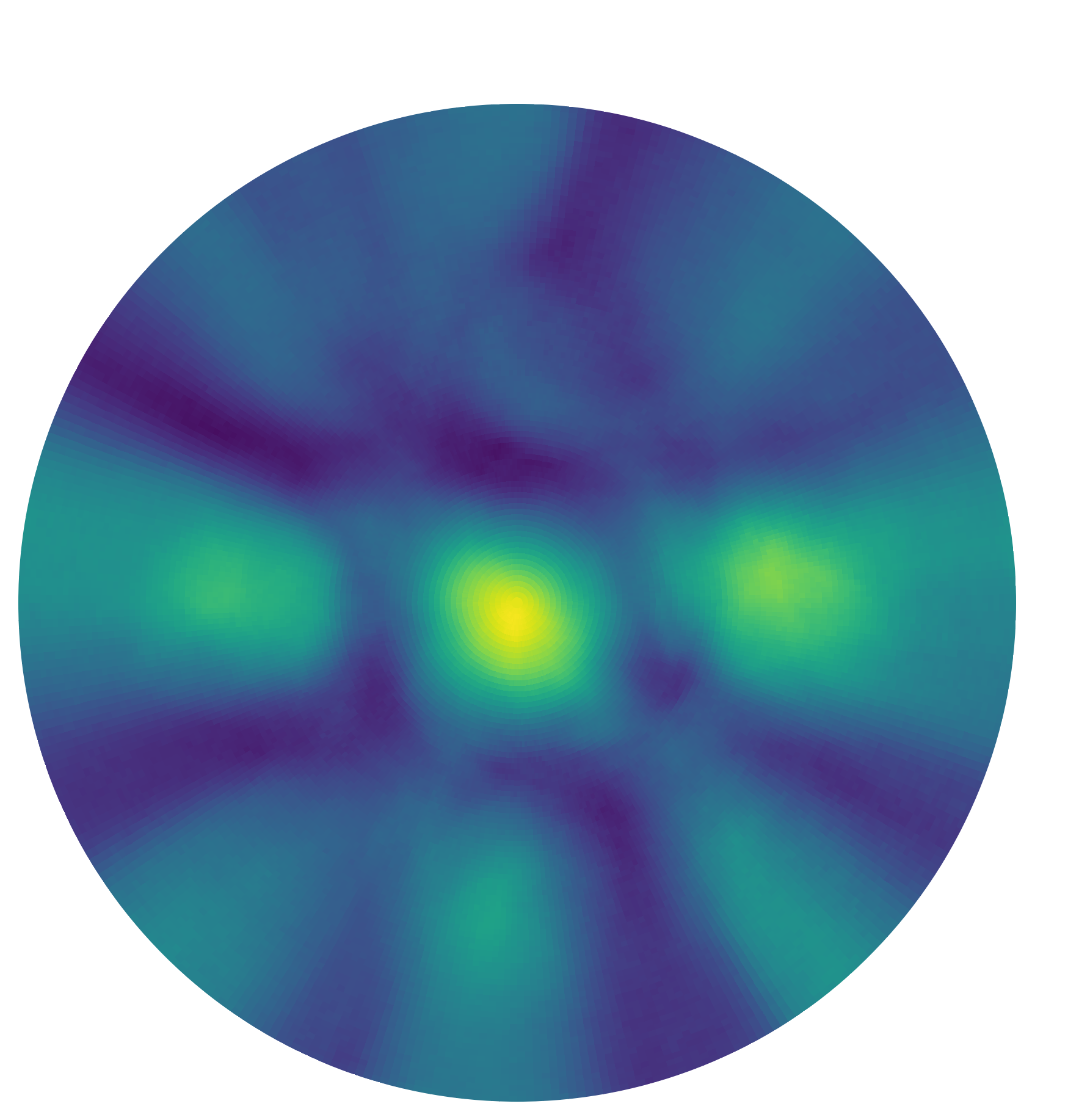} &
            \includegraphics[width=0.105\textwidth]{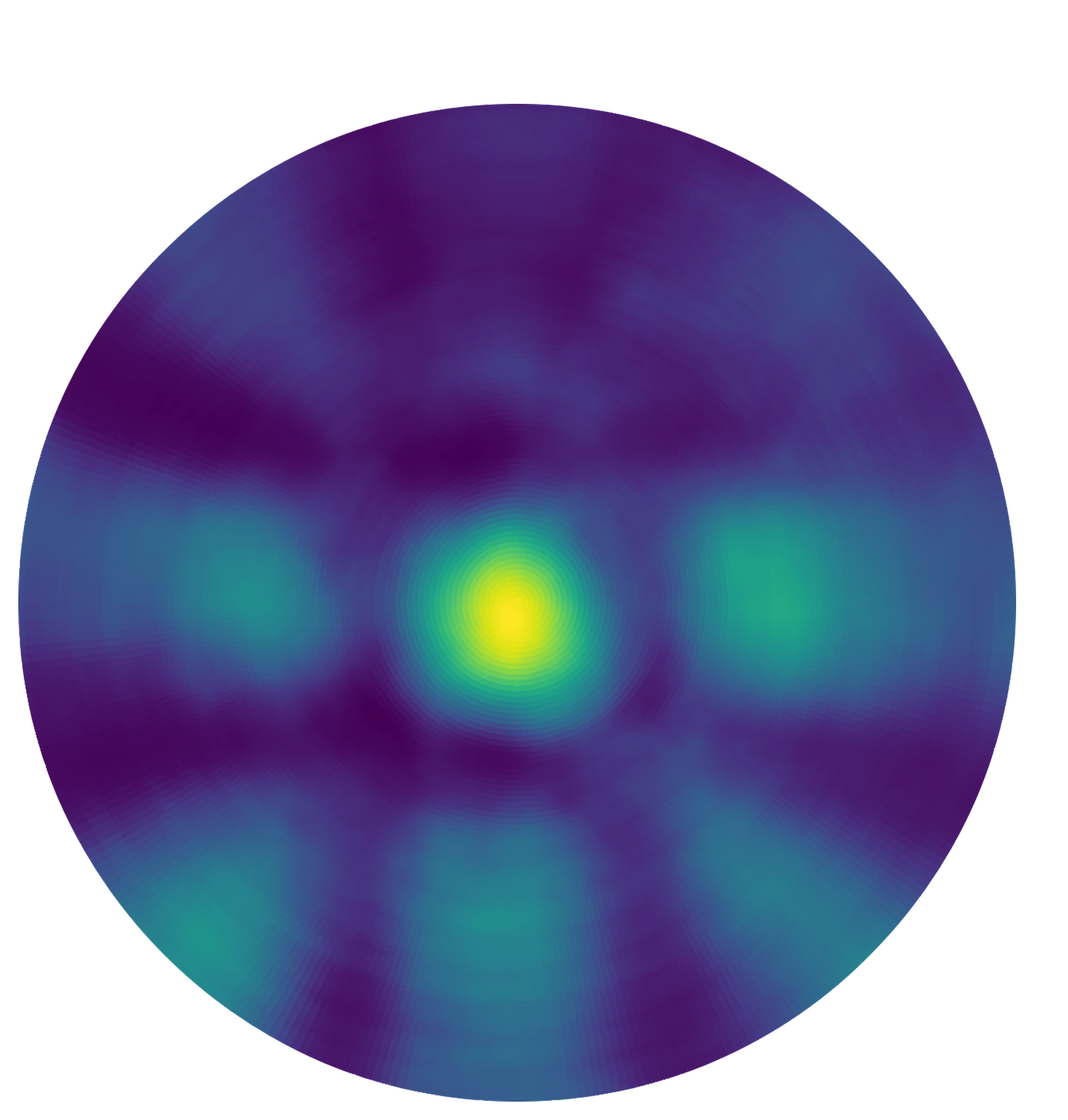} &
            \includegraphics[width=0.105\textwidth]{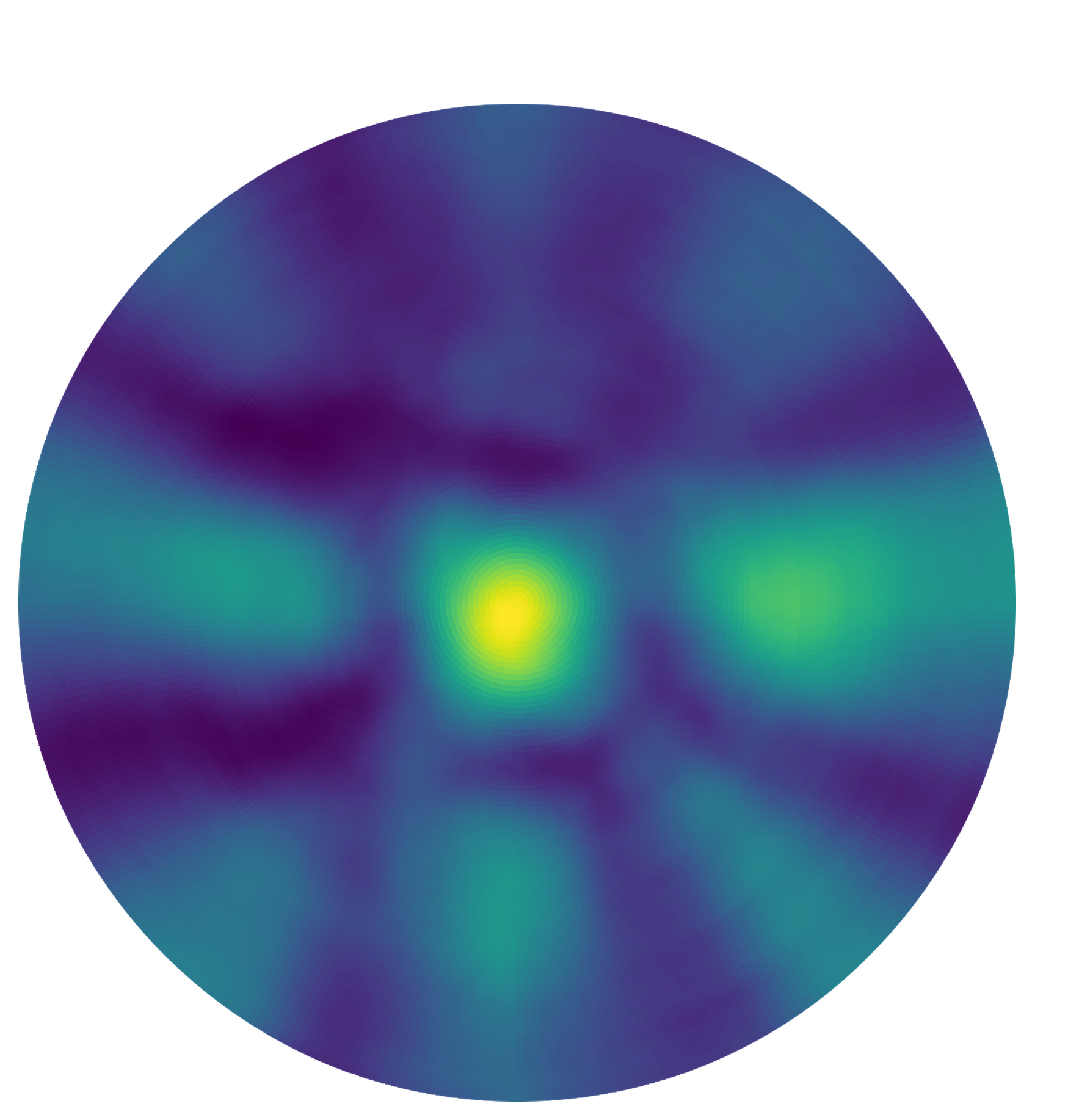} \\

            \raisebox{0.045\textwidth}{P4} &
            \includegraphics[width=0.105\textwidth]{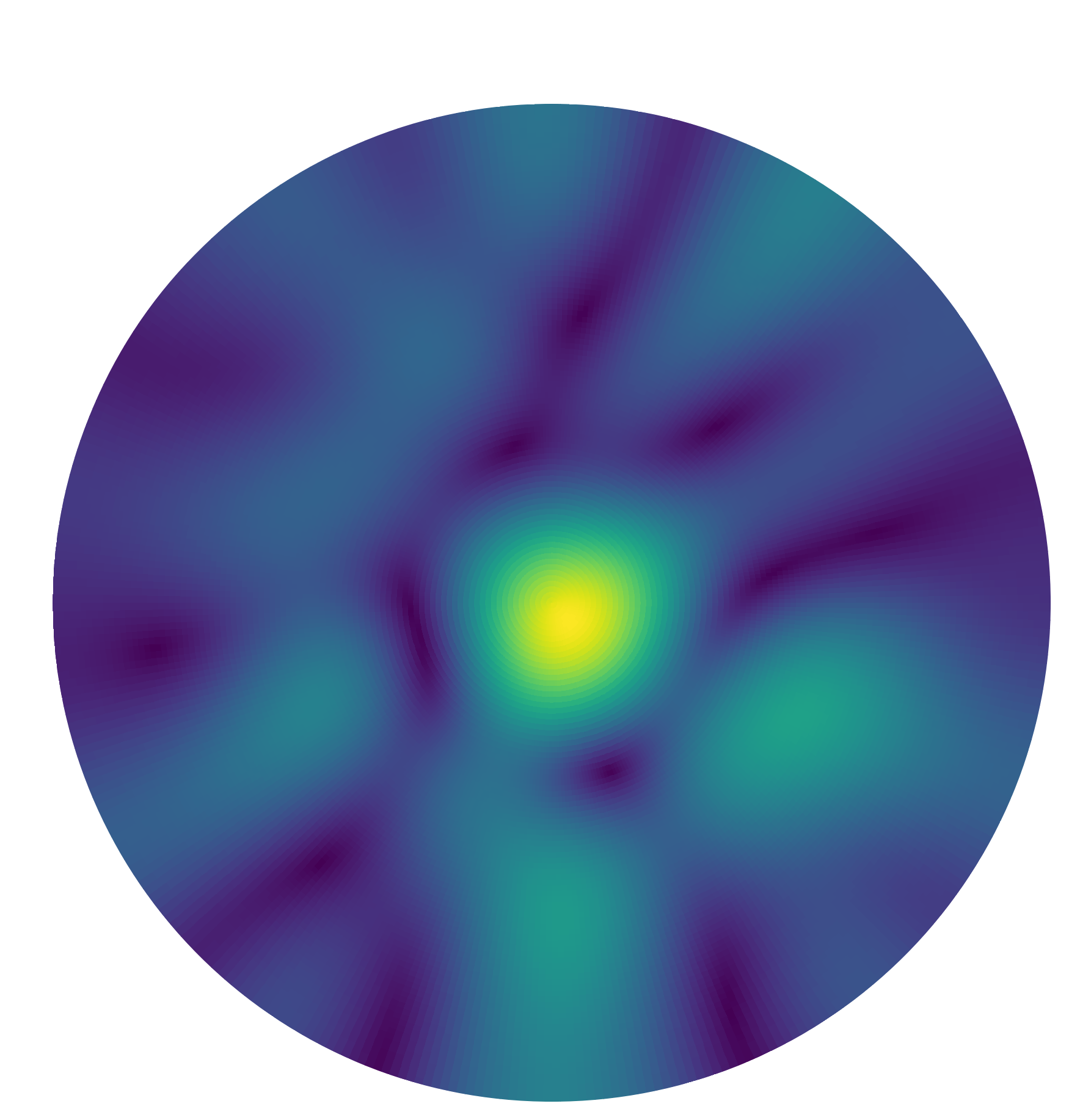} &
            \includegraphics[width=0.105\textwidth]{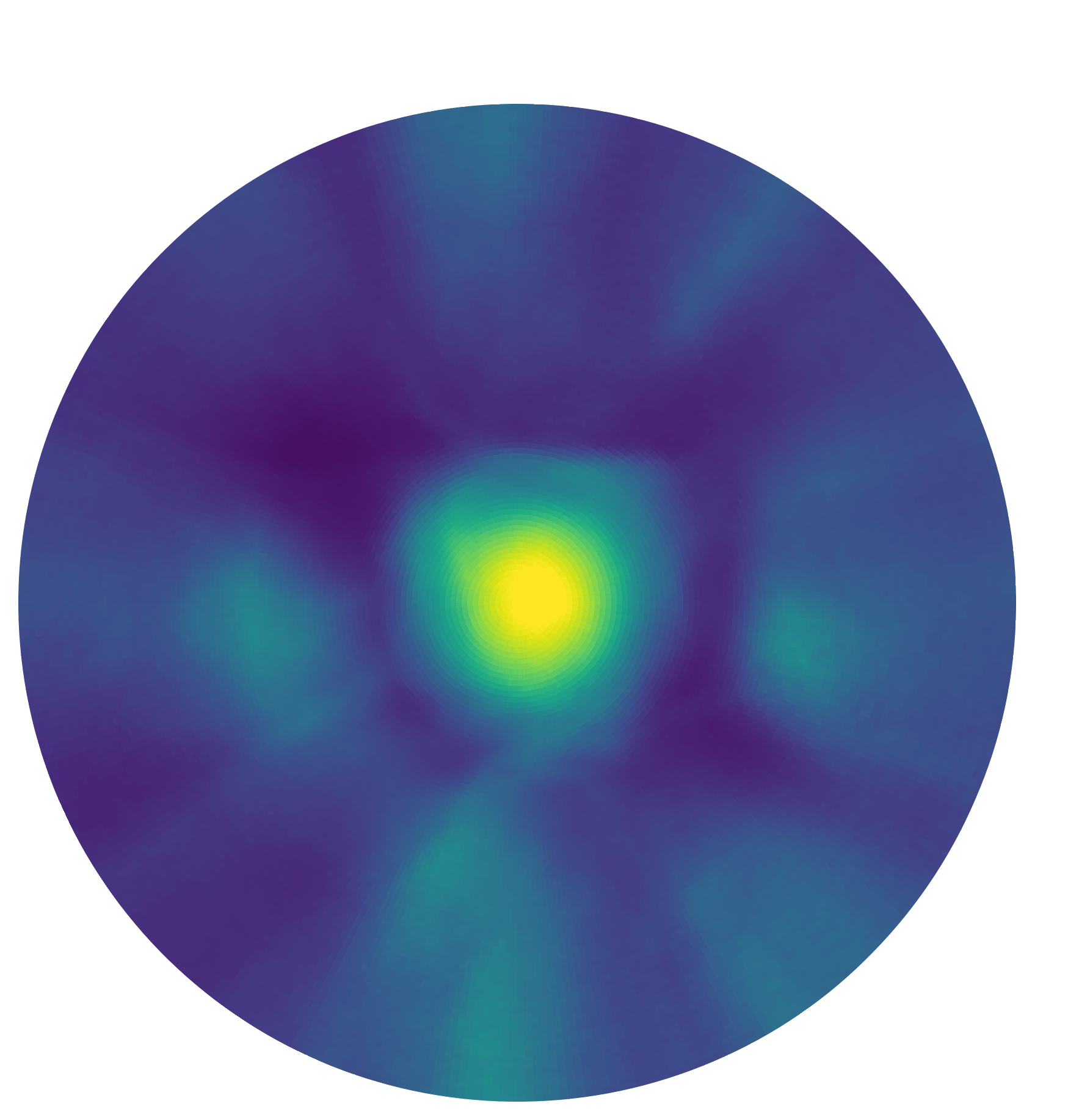} &
            \includegraphics[width=0.105\textwidth]{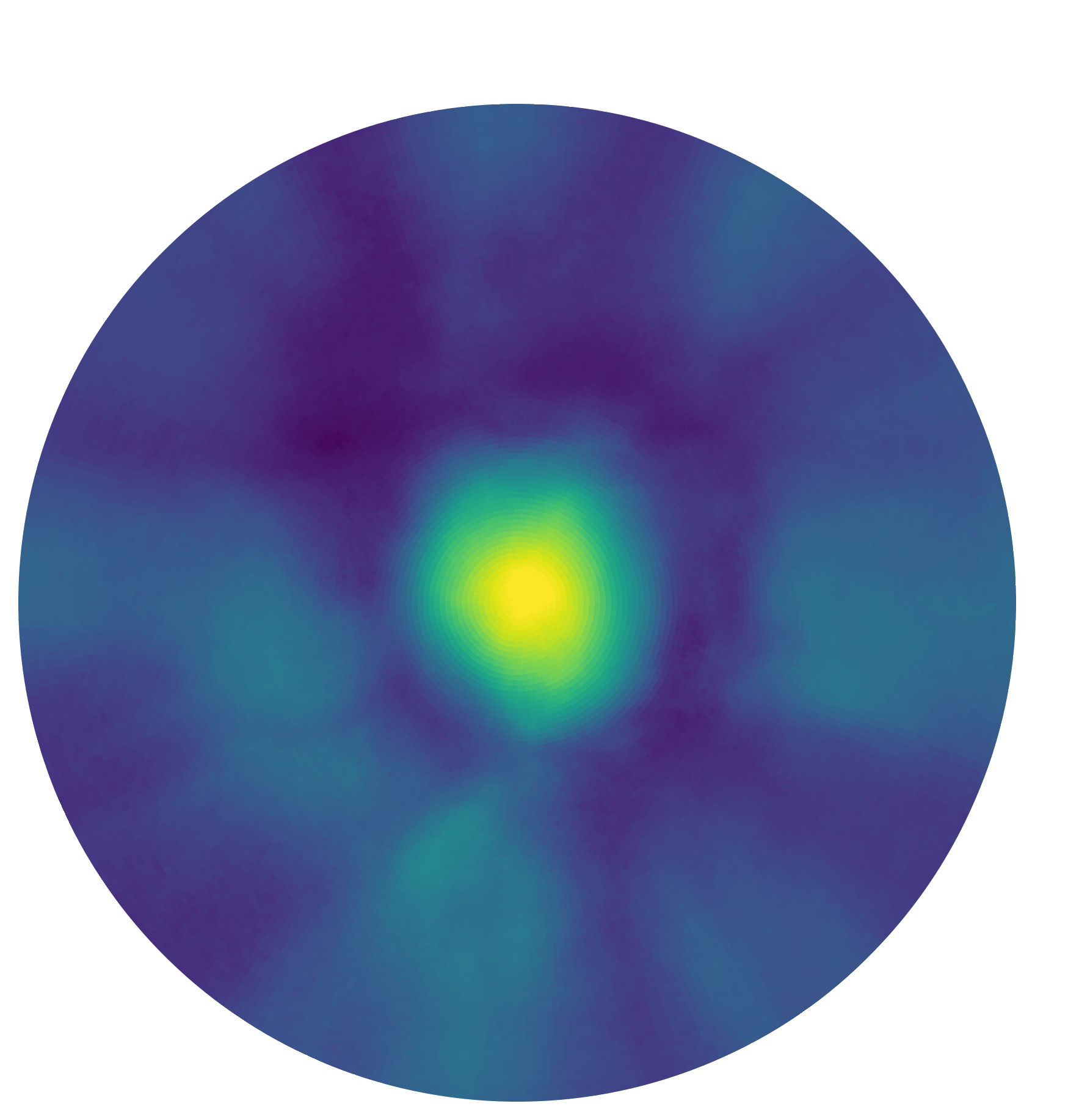} &
            \includegraphics[width=0.105\textwidth]{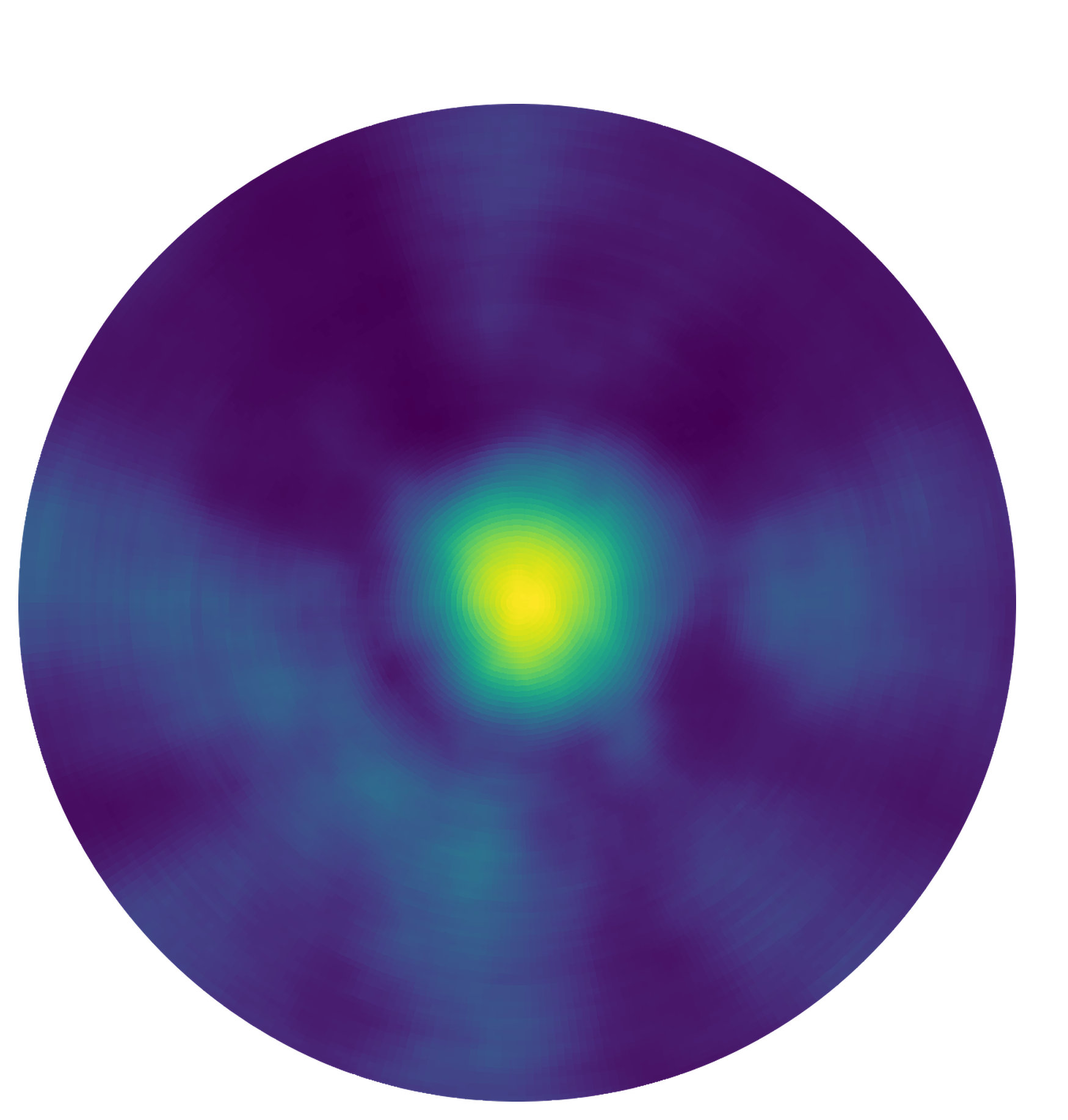} &
            \includegraphics[width=0.105\textwidth]{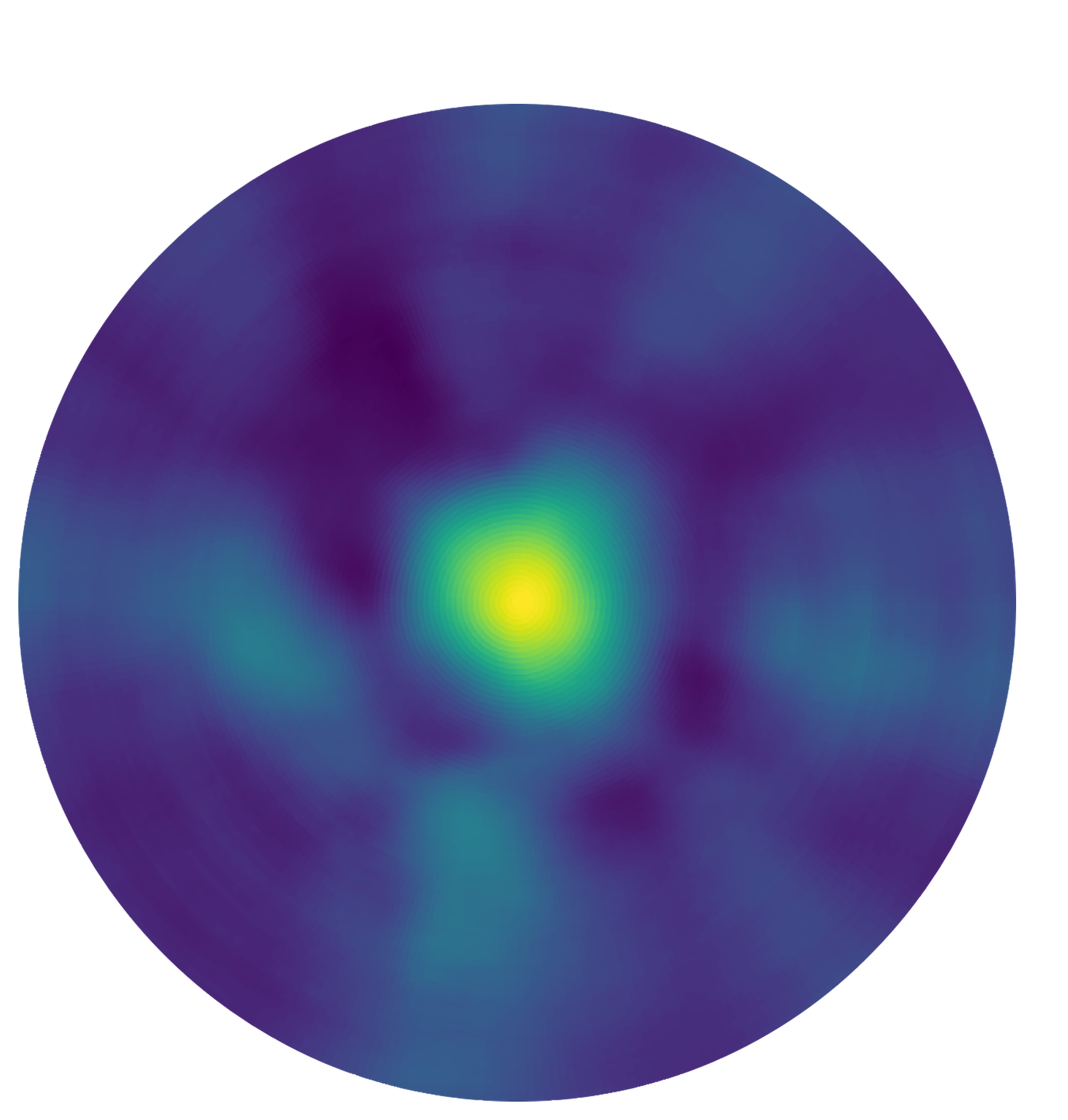} \\
            \bottomrule
        \end{tabular}
        \caption{Representative spatial-spectrum magnitude predictions. }
        \label{fig:realworld-spectrum-examples}
    \end{subfigure}

    \caption{Single-voxel performance on the real-world NeRF$^2$ dataset. SV-NeRF$^2$ and SV-INGP are compared against their multi-voxel baselines using SSIM and spatial-spectrum predictions shown across 4 TX locations (rows) and baselines (columns). Azimuth spand 0$^o$  to 360$^o$ around the plot; elevation spans 0$^o$  to 90$^o$ along the radius.}
    \label{fig:realworld-sv-results}
    \vspace{-6mm}
\end{figure*}

\subsection{Implementation Details}
\textbf{Datasets:}
We evaluate our single-voxel NeRF models on both experimental and simulated wireless datasets. For real-world evaluation, we use the public NeRF$^2$ dataset~\cite{nerf2}, collected in a cluttered indoor lab containing multiple reflectors like walls and objects that generate a complex dataset. It uses a fixed $4\times4$ Uniform Planar Array (UPA) RX that receives the signal backscattered from a mobile RFID tag. The dataset contains measured spatial spectra at 6,123 tag locations.

For generalization, we generate Sionna RT datasets~\cite{sionna} in a $3\,\mathrm{m}\times3\,\mathrm{m}$ enclosed indoor scene, with all walls made of metal for stronger multipath. The receiver uses an $8\times8$ UPA, and an omnidirectional TX is placed at multiple locations. Simulated channels are converted into ground-truth azimuth-elevation spatial spectra.


\textbf{Baselines:}
We compare against two multi-voxel wireless NeRF architectures: (i) NeRF$^2$ \cite{nerf2}, a wireless adaptation of Neural Radiance Fields, and (ii) Wireless INGP, our wireless implementation adapted from vision-based Instant Neural Graphics Primitives (INGP) \cite{ingp}. 

\textbf{Training Configuration:}
The NeRF$^2$ and Wireless INGP baselines use $64$ uniformly spaced samples per ray (as in ~\cite{nerf2}), whereas the proposed single-voxel NeRF variants use a single voxel per ray. The volume-density and radiance networks consist of one and two hidden layers, respectively, with $256$ neurons per layer. Models are trained for $30{,}000$ iterations using AdamW  optimizer~\cite{adam_optimizer} with learning rate $10^{-3}$ and batch size $8192$ rays. The implementation uses \texttt{tiny-cuda-nn} \cite{tiny-cuda-nn} and is trained on a single NVIDIA RTX A6000 GPU.

\textbf{Evaluation Metrics:}
Performance is evaluated using the structural similarity index measure (SSIM) between the predicted and ground-truth spatial spectra. We also report training and inference times to quantify computational complexity.

\subsection{Single-Voxel Wireless NeRF Evaluation}
We evaluate whether dense depth-wise sampling is necessary for spatial-spectrum prediction in a real RF environment. 
Fig.~\ref{fig:realworld-sv-results}(a) shows the CDF of SSIM across test transmitter locations. The single-voxel variants achieve SSIM distributions close to their corresponding multi-voxel baselines. In particular, SV-NeRF$^2$ remains comparable to NeRF$^2$ with only a minor 5.5\% reduction in SSIM, and SV-INGP closely tracks INGP with only a 2.5\% reduction in SSIM. Fig.~\ref{fig:realworld-sv-results}(b) shows the mean SSIM values, using the INGP model, for different numbers of voxels per ray. We observe that the mean SSIM remains flat across 1, 2, 4, 8, 16, and 32-voxel cases. This shows that dense depth-wise sampling is not necessary for spatial-spectrum prediction. Fig.~\ref{fig:realworld-sv-results}(c) shows representative spatial-spectrum predictions for four transmitter locations. The single-voxel models preserve the dominant spatial patterns and high-power regions observed in the ground truth, despite using only one voxel per ray. 


\begin{table}[!htbp]
\centering
\caption{Training and inference time for different models.}
\label{table:training-speed}
\small
\begin{tabular}{lccc}
\toprule
\textbf{Model} & \textbf{Median SSIM} &\textbf{Training time} & \textbf{Inference time}\\
\midrule
NeRF$^2$ & 0.78 & 7 h 40 min & 377 ms\\
INGP & 0.81 & 20 min & 48 ms\\
SV-NeRF$^2$& 0.74  & 1 h 40 min & 21 ms\\
SV-INGP & 0.78 & \textbf{2.5 min} & \textbf{4.7 ms}\\
\bottomrule
\end{tabular}
\end{table}
Table~\ref{table:training-speed} reports the training time of each model. NeRF$^2$ requires $7$ hours and $40$ minutes to train, while INGP reduces this time to $20$ minutes through multi-resolution hash encoding and CUDA acceleration. The proposed single-voxel variants further reduce training time. SV-NeRF$^2$ trains in $1$ hour and $40$ minutes, while SV-INGP trains in only $2.5$ minutes, 8x reduction compared to the multi-voxel INGP baseline. Notably, our SV-INGP model provides 184x reduction compared to the NeRF$^2$ baseline, while achieving the same SSIM of 0.78. Overall, these results show that dense depth-wise voxel sampling is not necessary to obtain comparable spatial-spectrum predictions on the real-world NeRF$^2$ dataset, while removing such sampling substantially reduces training cost.

\subsection{\textbf{Interpretability:} Understanding the Role of Volume Density}
Here, we first examine whether the learned volume density can be interpreted similarly to that in vision NeRFs. In vision NeRFs, the learned density behaves like an opacity or occupancy map, with large values occurring near visible surfaces in the scene. Wireless NeRFs use an analogous volume density $\sigma(P_x)$ for a voxel at position $P_x$, as defined in Eq.~(\ref{eq:single-voxel-ingp-power}).

\begin{figure}[!t]
\centering
\begin{subfigure}[t]{0.48\linewidth}
\centering
\includegraphics[width=\linewidth]{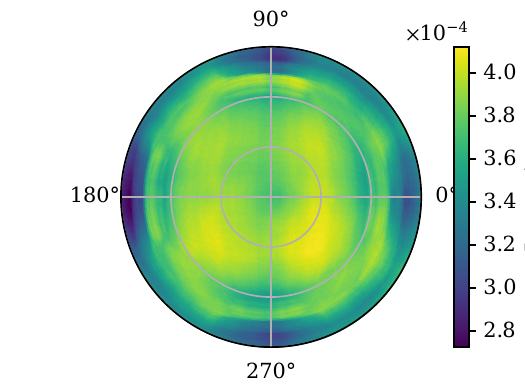}
\caption{Max volume density map}
\label{fig:nerf2-max-attn}
\end{subfigure}
\hfill
\begin{subfigure}[t]{0.48\linewidth}
\centering
\includegraphics[width=\linewidth]{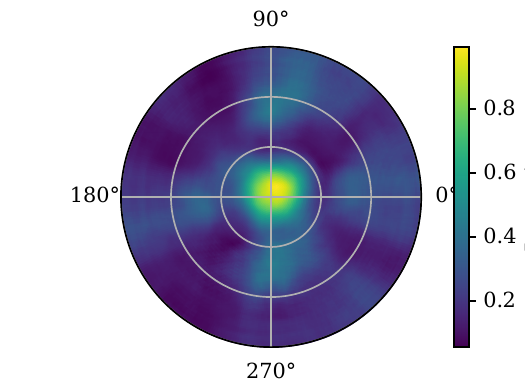}
\caption{Predicted spatial spectrum}
\label{fig:nerf2-pred-spectrum}
\end{subfigure}
\caption{Interpretability analysis: NeRF$^2$ maximum volume-density map does not clearly align with predicted spatial-spectrum peaks.}
\label{fig:nerf2-volume-density-and-pred-spatial-spectrum}
\vspace{-6mm}
\end{figure}

We study whether this learned density provides a similarly interpretable scene representation. For each angular ray, we compute the maximum volume density over all sampled voxels along that ray and compare this direction-wise density map with the predicted spatial spectrum. If the learned volume density directly encoded the RF scene structure, dominant received directions would be expected to show distinct density values. However, {Fig.~\ref{fig:nerf2-volume-density-and-pred-spatial-spectrum}(a) } shows that the maximum density map does not clearly align with the predicted spatial spectrum in Fig.~\ref{fig:nerf2-volume-density-and-pred-spatial-spectrum}(b). High-power spectrum directions are not consistently associated with large density values. \textit{This suggests that, unlike vision, the learned volume density is not directly interpretable as scene geometry for spatial-spectrum magnitude prediction.}

This behavior follows from the structure described in
Section~\ref{subsec:sv-insights}. The loss~\eqref{eq:loss-function} constrains only the overall signal strength~\eqref{eq:single-voxel-ingp-power}, but not the individual factors. Because the radiance network already receives $\omega$, it can reproduce any per-direction gain the density term would supply, so a range of density values fit the data equally well, and the split between the two is left to initialization and optimization. In vision NeRFs, the same scene point is constrained by rays from many camera poses, and this shared constraint is what forces density to settle on a geometrically meaningful value. A fixed receiver removes that constraint in the wireless counterpart, thus producing values that are not geometrically aligned with the scene structure.

\begin{figure}[!t]
\centering
\begin{subfigure}[t]{0.48\linewidth}
\centering
\includegraphics[width=\linewidth]{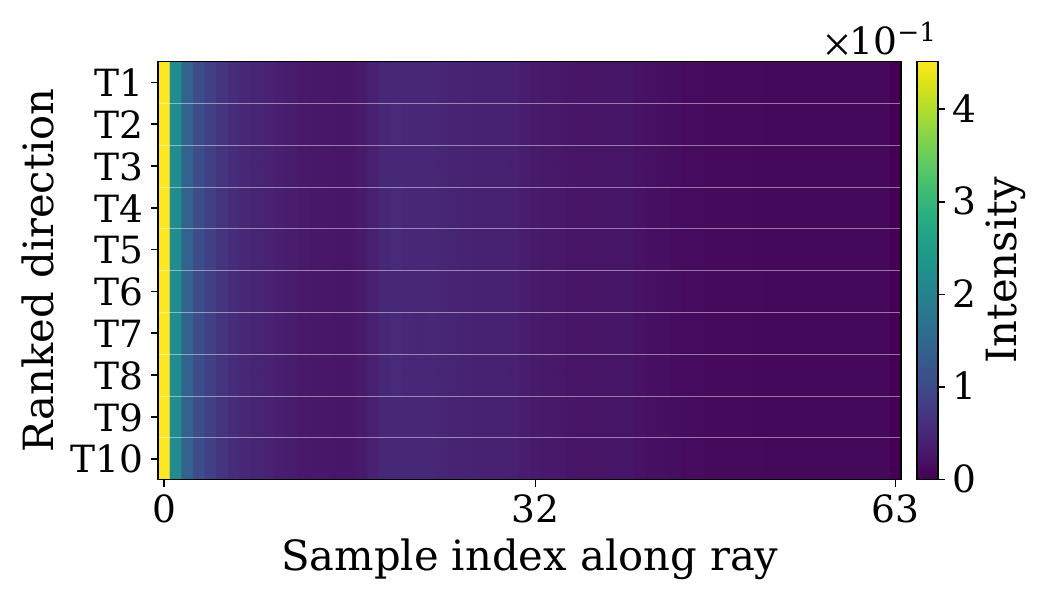}
\caption{NeRF$^2$ voxel contribution}
\label{fig:nerf2-top10-voxel}
\end{subfigure}
\hfill
\begin{subfigure}[t]{0.40\linewidth}
\centering
\includegraphics[width=\linewidth]{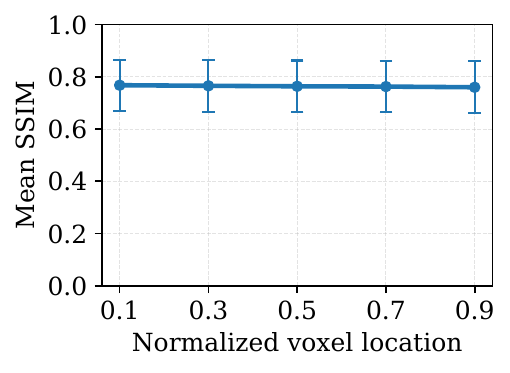}
\caption{Voxel placement impact}
\label{fig:ssim_vs_num_voxels}
\end{subfigure}
\caption{(a) Effective contribution of all the voxels along top-10 rays in multi-voxel NeRF$^2$ model (b) Impact of single voxel placement on end-to-end mean SSIM in SV-NeRF$^2$ model.}
\label{fig:top-10-volume-density-along-ray}
\vspace{-6mm}
\end{figure}

\subsection{\textbf{Interpretability:} Understanding Single Voxel Placement}
We next examine why dense voxel sampling can be reduced to a single voxel and where this voxel can be placed without substantially degrading spatial-spectrum prediction. We first train a 64-voxel NeRF$^2$ model and, for each voxel, compute its net contribution, including the learned volume density, the learned retransmitted signal, and the explicit path-loss factor. Fig.~\ref{fig:top-10-volume-density-along-ray}(a) shows the contribution along the strongest $10$ predicted rays. Although multiple voxels are sampled per ray, we found that most of the response is concentrated in the very first voxel.  When examining why it was the first voxel rather than any other on the ray, we found that the explicit path-loss factor yields the smallest attenuation for this first voxel because it is closest to the RX location. 

To verify whether voxel placement matters for the SV variant of NeRF, we implemented the SV-NeRF$^2$ model and varied the placement of a single voxel along the ray. Fig.~\ref{fig:top-10-volume-density-along-ray}(b) shows a flat SSIM response for all voxel placements (normalized to scene dimension), suggesting the voxel placement actually has no effect on end-to-end performance. \textit{This indicates that dense voxel sampling provides limited additional information for this task and motivates a randomly placed single-voxel design.}

\textbf{Why voxel placement does not matter?}
The analysis in Section~\ref{subsec:sv-insights} predicts this behavior. Each measurement is indexed by direction alone, so the model is asked to predict the signal arriving from a direction rather than the signal originating at a particular location along that direction. Any change in $l_{\rm sv}$ is absorbed by the radiance network, making the overall performance independent of voxel location. Fig.~\ref{fig:top-10-volume-density-along-ray}(b) confirms this empirically. The blocked (NLoS) case follows from the same reasoning. A blocker removes energy from a direction rather than from a depth, and since the measurement carries no depth information, the network assigns low intensity in that direction whether the voxel sits in front of or behind the blocker.

\subsection{Generalization Across Wireless Scenarios}

We next test whether the single-voxel finding holds beyond the real-world NeRF$^2$ dataset. For this study, we use the Sionna datasets and evaluate SV-INGP against its multi-voxel INGP baseline across different propagation settings, carrier frequencies, and array tapering configurations.

\begin{figure}[!htbp]
    \centering
    \begin{minipage}{0.48\linewidth}
        \centering
        \includegraphics[width=\linewidth]{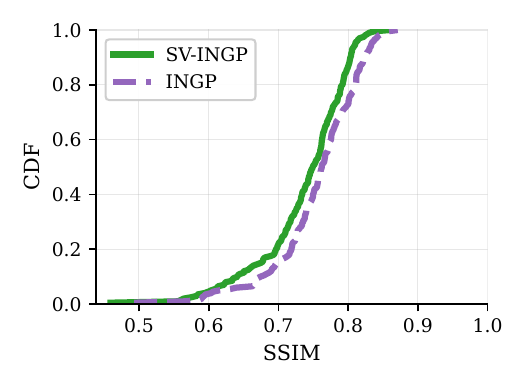}
        \small (a) LoS scene
    \end{minipage}
    \hfill
    \begin{minipage}{0.48\linewidth}
        \centering
        \includegraphics[width=\linewidth]{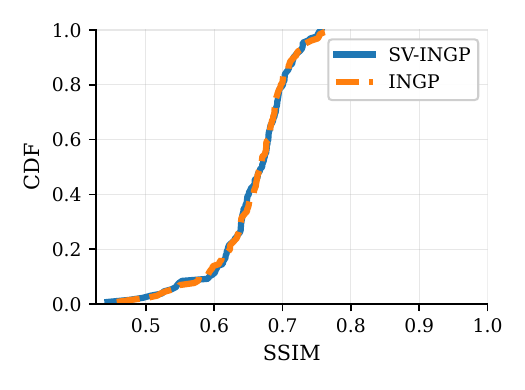}
        \small (b) NLoS scene
    \end{minipage}
    \caption{SSIM comparison in LoS and NLoS scenes at 27 GHz.}
    \label{fig:ingp-los-nlos-sv-mv}
    \vspace{-4mm}
\end{figure}

\textbf{LoS and NLoS scenes.}
The LoS scene (explained in section~\ref{sec:empirical evaluations}-A) contains a direct path and weaker reflected paths. For the NLoS scene, we place an obstacle (a plane sheet of metal) between the transmitter and receiver to attenuate the direct path while preserving reflected multipath components. Fig.~\ref{fig:ingp-los-nlos-sv-mv} shows that both models achieve lower SSIM in the NLoS setting, as expected. \textit{Still, the gap between SV-INGP and INGP remains small in both cases, indicating that multi-voxel sampling provides limited benefit even when the direct path is blocked.}

\begin{figure}[!htbp]
    \centering
    \begin{minipage}{0.48\linewidth}
        \centering
        \includegraphics[width=\linewidth]{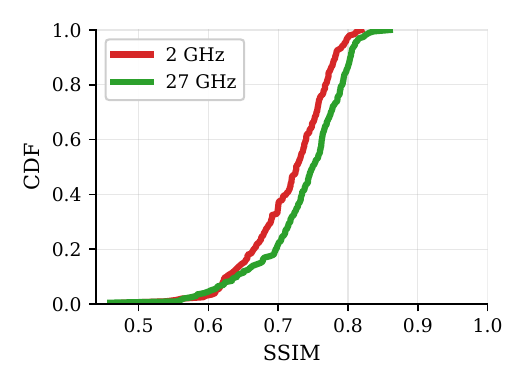}
        \small (a) 2.4 GHz versus 27 GHz
    \end{minipage}
    \hfill
    \begin{minipage}{0.48\linewidth}
        \centering
        \includegraphics[width=\linewidth]{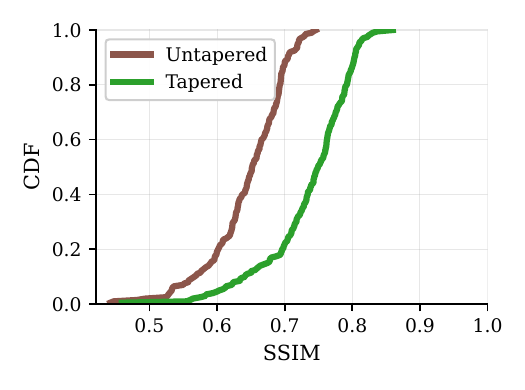}
        \small (b) Effect of tapering
    \end{minipage}
    \caption{SSIM across carrier frequency and array tapering.}
    \label{fig:ingp-multi-frequency-and-tapering}
    \vspace{-4mm}
\end{figure}

\textbf{Carrier frequency.}
We then evaluate SV-INGP at $2.4$ GHz and $27$ GHz using the same scene geometry, model architecture, and ray-sampling configuration. Fig.~\ref{fig:ingp-multi-frequency-and-tapering}(a) shows that the SSIM distributions remain close across the two frequencies. \textit{This suggests that moving to a millimeter-wave carrier does not necessarily require increasing the number of voxels per ray for spatial-spectrum magnitude prediction in this setting.}

\textbf{Array tapering.}
Finally, we study the effect of receiver-array tapering. Untapered arrays can create strong sidelobes in the spatial-spectrum target, which introduce angular artifacts that are not due to separate propagation paths. Fig.~\ref{fig:ingp-multi-frequency-and-tapering}(b) shows that \textit{tapering improves SSIM by suppressing these sidelobes and producing a cleaner prediction target with the usual tradeoff of a wider main lobe}.

\raggedbottom
\section{Limitations and Future Work}
Although the single-voxel NeRF results are promising, several limitations remain. First,
training on locally normalized spectra captures the relative power
distribution across the angular domain at each TX location but discards
relative power levels across TX locations, which matters for applications such
as path planning. Second, we consider only static scenes, so robustness to
environmental change is untested. Third, our experiments are limited to indoor environments, and extending the framework outdoors remains an important direction. Finally, these models are assessed only as an image reconstruction task. How the predicted spectra perform in actual wireless applications remains to be studied; using them to drive data-driven applications on platforms such as Open-RAN using RIC-based platforms~\cite{mimo-ric}, for tasks like beam management and localization, is left to future work.


\section{Conclusion}
\label{sec:conclusion}
This paper shows that dense volumetric sampling is not always necessary for
predicting wireless spatial-spectrum magnitudes. Our single-voxel design
achieves a median SSIM of $0.78$, matching the NeRF$^2$ baseline, while
reducing training time by 184x. We attribute this to the fixed-receiver geometry of the task: each measurement is indexed by direction alone, which leaves depth unconstrained and makes the learned volume density uninterpretable as scene structure. Current wireless NeRF models are therefore over-parameterized for this prediction task. Whether the same holds for representations that do not rely on volumetric density, such as Gaussian Splatting, is an open question.


\noindent
\textbf{Acknowledgment:}
This work was supported in part by the RPI-IBM Future Computing Research Center. We thank Alberto Valdes-Garcia and Alexandra Gallyas-Sanhueza from IBM for valuable feedback and discussions. We also thank RPI WAYS Lab members for constructive discussions.

\bibliographystyle{IEEEtran}
\bibliography{references}

@misc{deepmimo,
      title={{DeepMIMO: A Generic Deep Learning Dataset for Millimeter Wave and Massive MIMO Applications}}, 
      author={Ahmed Alkhateeb},
      howpublished={arXiv:1902.06435},
      year={2019},
      eprint={1902.06435},
      archivePrefix={arXiv},
      primaryClass={cs.IT},
      url={https://arxiv.org/abs/1902.06435}, 
}

@article{deepsense_6g,
  author  = {Alkhateeb, Ahmed and Charan, Gouranga and Osman, Tawfik and Hredzak, Andrew
             and Morais, Joao and Demirhan, Umut and Srinivas, Nikhil},
  title   = {{DeepSense 6G: A Large-Scale Real-World Multi-Modal Sensing and Communication Dataset}},
  journal = {IEEE Communications Magazine},
  volume  = {61},
  number  = {9},
  pages   = {122--128},
  year    = {2023},
  doi     = {10.1109/MCOM.006.2200730}
}

@inproceedings{csi_benchmark_dataset,
  author    = {Zhu, Guozhen and Hu, Yuqian and Gao, Weihang
               and Wang, Wei-Hsiang and Wang, Beibei and Liu, K. J. Ray},
  title     = {{CSI-Bench: A Large-Scale In-the-Wild Dataset for Multi-task WiFi Sensing}},
  booktitle = {Proceedings of Advances in Neural Information Processing Systems (NeurIPS)},
  address   = {San Diego, CA},
  volume    = {38},
  month     = dec,
  year      = {2025},
  doi       = {10.52202/085713-5648}
}

@article{digital_twin_6g,
  author  = {Tao, Zhenyu and Xu, Wei and Huang, Yongming and Wang, Xiaoyun and You, Xiaohu},
  title   = {{Wireless network digital twin for 6G: Generative AI as a key enabler}},
  journal = {IEEE Wireless Communications},
  volume  = {31},
  number  = {4},
  pages   = {24--31},
  year    = {2024}
}

@article{sionna_ray_tracing_hyodis,
  author  = {Hoydis, Jakob and A{\"i}t Aoudia, Fay{\c c}al and Cammerer, Sebastian
             and Euchner, Florian and Nimier-David, Merlin and ten Brink, Stephan
             and Keller, Alexander},
  title   = {{Learning Radio Environments by Differentiable Ray Tracing}},
  journal = {IEEE Transactions on Machine Learning in Communications and Networking},
  volume  = {2},
  pages   = {1527--1539},
  year    = {2024},
  doi     = {10.1109/TMLCN.2024.3474639}
}

@article{pilot_cavers,
  author  = {Cavers, James K.},
  title   = {{An analysis of pilot symbol assisted modulation for Rayleigh fading channels (mobile radio)}},
  journal = {IEEE Transactions on Vehicular Technology},
  volume  = {40},
  number  = {4},
  pages   = {686--693},
  year    = {1991},
  doi     = {10.1109/25.108378}
}

@inproceedings{pilot_ofdm_van_de_beek,
  author    = {{van de Beek}, Jan-Jaap and Edfors, Ove and Sandell, Magnus
               and Wilson, Sarah Kate and B{\"o}rjesson, Per Ola},
  title     = {On channel estimation in {OFDM} systems},
  booktitle = {Proceedings of the IEEE Vehicular Technology Conference (VTC)},
  address   = {Chicago, IL},
  volume    = {2},
  pages     = {815--819},
  month     = jul,
  year      = {1995},
  doi       = {10.1109/VETEC.1995.504981}
}

@article{dl_channel_estimation_and_signal_detection_hao_2018,
  author  = {Ye, Hao and Li, Geoffrey Ye and Juang, Biing-Hwang},
  title   = {{Power of Deep Learning for Channel Estimation and Signal Detection in OFDM Systems}},
  journal = {IEEE Wireless Communications Letters},
  volume  = {7},
  number  = {1},
  pages   = {114--117},
  year    = {2018},
  doi     = {10.1109/LWC.2017.2757490}
}

@article{dl_based_channel_estimation_soltani_2019,
  author  = {Soltani, Mehran and Pourahmadi, Vahid and Mirzaei, Ali and Sheikhzadeh, Hamid},
  title   = {{Deep Learning-Based Channel Estimation}},
  journal = {IEEE Communications Letters},
  volume  = {23},
  number  = {4},
  pages   = {652--655},
  year    = {2019},
  doi     = {10.1109/LCOMM.2019.2898944}
}

@article{dl_channel_estimation_beamspace_mmwave_mimo_hengtao_2018,
  author  = {He, Hengtao and Wen, Chao-Kai and Jin, Shi and Li, Geoffrey Ye},
  title   = {{Deep Learning-Based Channel Estimation for Beamspace mmWave Massive MIMO Systems}},
  journal = {IEEE Wireless Communications Letters},
  volume  = {7},
  number  = {5},
  pages   = {852--855},
  month   = oct,
  year    = {2018},
  doi     = {10.1109/LWC.2018.2832128}
}

@article{ai_ml_for_beam_management,
  author  = {Xue, Qing and Guo, Jiajia and Zhou, Binggui and Xu, Yongjun and Li, Zhidu and Ma, Shaodan},
  title   = {{AI/ML for Beam Management in 5G-Advanced: A Standardization Perspective}},
  journal = {IEEE Vehicular Technology Magazine},
  volume  = {19},
  number  = {4},
  pages   = {64--72},
  year    = {2024},
  doi     = {10.1109/MVT.2024.3431790}
}

@ARTICLE{beam_management_for_dense_mmwave_network,
  author={Xue, Qing and Liu, Yi-Jing and Sun, Yao and Wang, Jian and Yan, Li and Feng, Gang and Ma, Shaodan},
  journal={IEEE Transactions on Cognitive Communications and Networking}, 
  title={{Beam Management in Ultra-Dense mmWave Network via Federated Reinforcement Learning: An Intelligent and Secure Approach}}, 
  year={2023},
  volume={9},
  number={1},
  pages={185--197},
  doi={10.1109/TCCN.2022.3215527}}

@inproceedings{two_beams_2021,
  author    = {Jain, Ish Kumar and Subbaraman, Raghav and Bharadia, Dinesh},
  title     = {Two beams are better than one: Towards reliable and high throughput {mmWave} links},
  booktitle = {Proceedings of the ACM SIGCOMM Conference},
  address   = {Virtual event},
  pages     = {488--502},
  month     = aug,
  year      = {2021},
  doi       = {10.1145/3452296.3472924}
}

@ARTICLE{isac_survey,
  author={Liu, An and Huang, Zhe and Li, Min and Wan, Yubo and Li, Wenrui and Han, Tony Xiao and Liu, Chenchen and Du, Rui and Tan, Danny Kai Pin and Lu, Jianmin and Shen, Yuan and Colone, Fabiola and Chetty, Kevin},
  journal={IEEE Communications Surveys \& Tutorials}, 
  title={{A Survey on Fundamental Limits of Integrated Sensing and Communication}}, 
  year={2022},
  volume={24},
  number={2},
  pages={994--1034},
  doi={10.1109/COMST.2022.3149272}}

@ARTICLE{isac_survey_2026,
  author={Zhang, Di and Cui, Yuanhao and Cao, Xiaowen and Su, Nanchi and Gong, Yi and Liu, Fan and Yuan, Weijie and Jing, Xiaojun and Zhang, J. Andrew and Xu, Jie and Masouros, Christos and Niyato, Dusit and Di Renzo, Marco},
  journal={IEEE Communications Surveys \& Tutorials}, 
  title={{Integrated Sensing and Communications Over the Years: An Evolution Perspective}}, 
  year={2026},
  volume={28},
  number={},
  pages={5014--5048},
  doi={10.1109/COMST.2026.3655674}}

@ARTICLE{localization_2015,
  author={Van Nguyen, Thang and Jeong, Youngmin and Shin, Hyundong and Win, Moe Z.},
  journal={IEEE Journal on Selected Areas in Communications}, 
  title={{Machine Learning for Wideband Localization}}, 
  year={2015},
  volume={33},
  number={7},
  pages={1357--1380},
  doi={10.1109/JSAC.2015.2430191}}

@ARTICLE{localization_metaloc_2023,
  author={Gao, Jun and Wu, Dongze and Yin, Feng and Kong, Qinglei and Xu, Lexi and Cui, Shuguang},
  journal={IEEE Journal on Selected Areas in Communications}, 
  title={{MetaLoc: Learning to Learn Wireless Localization}}, 
  year={2023},
  volume={41},
  number={12},
  pages={3831--3847},
  doi={10.1109/JSAC.2023.3322766}}

@article{nerf,
  author  = {Mildenhall, Ben and Srinivasan, Pratul P. and Tancik, Matthew and Barron, Jonathan T. and Ramamoorthi, Ravi and Ng, Ren},
  title   = {{NeRF: Representing scenes as neural radiance fields for view synthesis}},
  journal = {Communications of the ACM},
  volume  = {65},
  number  = {1},
  pages   = {99--106},
  month   = jan,
  year    = {2022},
  doi     = {10.1145/3503250}
}

@inproceedings{mip_nerf,
  title={{Mip-NeRF: A Multiscale Representation for Anti-Aliasing Neural Radiance Fields}},
  author={Barron, Jonathan T and Mildenhall, Ben and Tancik, Matthew and Hedman, Peter and Martin-Brualla, Ricardo and Srinivasan, Pratul P},
  booktitle={Proceedings of the IEEE/CVF International Conference on Computer Vision (ICCV)},
  pages={5835--5844},
  year={2021},
  organization={IEEE}
}

@inproceedings{nerf_wo_camera_neurips,
  author    = {Amballa, Chaitanya and Wei, Yu-Lin and Basu, Sattwik
               and Yang, Zhijian and Ergezer, Mehmet and Roy Choudhury, Romit},
  title     = {{Can NeRFs ``See'' without Cameras?}},
  booktitle = {Proceedings of Advances in Neural Information Processing Systems (NeurIPS)},
  address   = {San Diego, CA},
  volume    = {38},
  pages     = {112023--112050},
  month     = dec,
  year      = {2025},
  doi       = {10.52202/085713-3743}
}

@inproceedings{dart_radar_nerf,
  author    = {Huang, Tianshu and Miller, John and Prabhakara, Akarsh and Jin, Tao
               and Laroia, Tarana and Kolter, Zico and Rowe, Anthony},
  title     = {{DART: Implicit Doppler tomography for radar novel view synthesis}},
  booktitle = {Proceedings of the IEEE/CVF Conference on Computer Vision and Pattern Recognition (CVPR)},
  address   = {Seattle, WA},
  pages     = {24118--24129},
  month     = jun,
  year      = {2024}
}

@inproceedings{nerf2,
  author    = {Zhao, Xiaopeng and An, Zhenlin and Pan, Qingrui and Yang, Lei},
  title     = {{NeRF$^{2}$: Neural Radio-Frequency Radiance Fields}},
  booktitle = {Proceedings of the 29th Annual International Conference on Mobile Computing and Networking (MobiCom)},
  address   = {Madrid, Spain},
  month     = oct,
  year      = {2023},
  note      = {{Art. no.}~27},
  doi       = {10.1145/3570361.3592527}
}

@article{gaussian_splat_cv,
  author  = {Kerbl, Bernhard and Kopanas, Georgios and Leimk{\"u}hler, Thomas and Drettakis, George},
  title   = {{3D Gaussian splatting for real-time radiance field rendering}},
  journal = {ACM Transactions on Graphics},
  volume  = {42},
  number  = {4},
  month   = jul,
  year    = {2023},
  note    = {{Art. no.}~139},
  doi     = {10.1145/3592433}
}

@inproceedings{wrf_gs_wen_2025,
  author    = {Wen, Chaozheng and Tong, Jingwen and Hu, Yingdong and Lin, Zehong and Zhang, Jun},
  title     = {{WRF-GS: Wireless Radiation Field Reconstruction with 3D Gaussian Splatting}},
  booktitle = {Proceedings of the IEEE Conference on Computer Communications (INFOCOM)},
  address   = {London, United Kingdom},
  pages     = {1--10},
  month     = may,
  year      = {2025},
  doi       = {10.1109/INFOCOM55648.2025.11044513}
}

@article{gsparc_tamu,
  title={{GSpaRC: Gaussian Splatting for Real-time Reconstruction of RF Channels}},
  author={Nukapotula, Bhavya Sai and Tripathi, Rishabh and Pregler, Seth and Kalathil, Dileep and Shakkottai, Srinivas and Rappaport, Theodore S},
  journal={arXiv preprint arXiv:2511.22793},
  year={2025}
}

@inproceedings{newrf,
  author    = {Lu, Haofan and Vattheuer, Christopher and Mirzasoleiman, Baharan and Abari, Omid},
  title     = {{NeWRF: A deep learning framework for wireless radiation field reconstruction and channel prediction}},
  booktitle = {Proceedings of the 41st International Conference on Machine Learning (ICML)},
  series    = {PMLR},
  address   = {Vienna, Austria},
  volume    = {235},
  pages     = {33147--33159},
  month     = jul,
  year      = {2024}
}

@article{ingp,
  author  = {M{\"u}ller, Thomas and Evans, Alex and Schied, Christoph and Keller, Alexander},
  title   = {{Instant neural graphics primitives with a multiresolution hash encoding}},
  journal = {ACM Transactions on Graphics},
  volume  = {41},
  number  = {4},
  month   = jul,
  year    = {2022},
  note    = {{Art. no.}~102},
  doi     = {10.1145/3528223.3530127}
}

@article{voxelrf,
  title={{VoxelRF: Voxelized radiance field for fast wireless channel modeling}},
  author={Zeng, Zihang and Sun, Shu and Tao, Meixia and Xu, Yin and Yu, Xianghao},
  journal={IEEE Communications Letters},
  volume={30},
  pages={617--621},
  year={2025},
  publisher={IEEE}
}

@inproceedings{nerf-apt,
  author    = {Shen, Jingzhou and Zhao, Tianya and Wu, Yanzhao and Wang, Xuyu},
  title     = {{NeRF-APT: A New NeRF Framework for Wireless Channel Prediction}},
  booktitle = {Proceedings of the IEEE Conference on Computer Communications Workshops (INFOCOM WKSHPS)},
  address   = {London, United Kingdom},
  pages     = {1--6},
  month     = may,
  year      = {2025},
  doi       = {10.1109/INFOCOMWKSHPS65812.2025.11152993}
}

@article{gs_rf-3dgs_60GHz,
  title={{RF-3DGS: Wireless channel modeling with radio radiance field and 3D Gaussian splatting}},
  author={Zhang, Lihao and Sun, Haijian and Berweger, Samuel and Gentile, Camillo and Hu, Rose Qingyang},
  journal={IEEE Transactions on Wireless Communications},
  volume={25},
  pages={10419--10433},
  year={2026},
  publisher={IEEE}
}

@ARTICLE{nerf2-ss,
  author={Liu, Kaicheng and Jiang, Wenjun and Yuan, Xiaojun},
  journal={IEEE Wireless Communications Letters}, 
  title={{Scene Structure Based Neural Radio-Frequency Radiance Fields for Channel Knowledge Map Construction}}, 
  year={2026},
  volume={15},
  number={},
  pages={171-175},
  doi={10.1109/LWC.2025.3615621}}

@inproceedings{rfcanvas,
  author    = {Chen, Xingyu and Feng, Zihao and Sun, Ke and Qian, Kun and Zhang, Xinyu},
  title     = {{RFCanvas: Modeling RF Channel by Fusing Visual Priors and Few-shot RF Measurements}},
  booktitle = {Proceedings of the 22nd ACM Conference on Embedded Networked Sensor Systems (SenSys)},
  address   = {Hangzhou, China},
  pages     = {464--477},
  month     = nov,
  year      = {2024},
  doi       = {10.1145/3666025.3699351}
}

@software{sionna,
 title = {{Sionna}},
 author = {Hoydis, Jakob and Cammerer, Sebastian and {Ait Aoudia}, Fayçal and Nimier-David, Merlin and Maggi, Lorenzo and Marcus, Guillermo and Vem, Avinash and Keller, Alexander},
 note = {https://nvlabs.github.io/sionna/},
 year = {2022},
 version = {2.1.0}
}

@article{adam_optimizer,
  title={{Decoupled Weight Decay Regularization}},
  author={Loshchilov, Ilya and Hutter, Frank},
  journal={arXiv preprint arXiv:1711.05101},
  year={2017}
}

@misc{tiny-cuda-nn,
  author       = {Thomas M{\"u}ller},
  title        = {{tiny-cuda-nn}},
  year         = {2021},
  month        = apr,
  version      = {2.0},
  howpublished = {\url{https://github.com/NVlabs/tiny-cuda-nn}},
  note         = {{BSD-3-Clause} license}
}

@inproceedings{mimo-ric,
  author    = {Jonnavithula, Sesha Sai Rakesh and Jain, Ish Kumar and Bharadia, Dinesh},
  title     = {{MIMO-RIC: RAN Intelligent Controller for MIMO xApps}},
  booktitle = {Proceedings of the 30th Annual International Conference on Mobile Computing and Networking (MobiCom)},
  address   = {Washington, DC},
  pages     = {2315--2322},
  year      = {2024},
  doi       = {10.1145/3636534.3701548}
}

\end{document}